\documentclass[a4paper,fleqn,usenatbib]{mnras}
\usepackage[T1]{fontenc}
\usepackage{ae,aecompl}
\usepackage{amsmath}
\usepackage{amsfonts}
\usepackage{amssymb}
\usepackage{graphicx}
\usepackage{color,xcolor}
\def\mnras{MNRAS}
\def\aap{A$\&$A}
\def\nat{Nature}
\def\apj{ApJ}
\def\apjl{ApJ Letters}
\def\apjs{ApJS}

\def\aj{AJ}

\newcommand{\me}{\, {\rm M}_{\oplus}}

\newcommand{\au}{\, {\rm au}}
\newcommand{\ab}{\, a_{\rm b}}

\newcommand{\dtg}{{\rm dust-to-gas~}}
\newcommand{\Stokes}{{St}}
\title[Dusty Circumbinary Discs]{Dusty circumbinary discs: inner cavity structures and stopping locations of migrating planets}
\author[G. A. L. Coleman et al]{Gavin A. L. Coleman$^1$\thanks{Email: gavin.coleman@qmul.ac.uk}, Richard. P. Nelson$^1$, Amaury H. M. J. Triaud$^2$\\
1. Astronomy Unit, Queen Mary University of London, Mile End Road, London, E1 4NS, United Kingdom\\
2. School of Physics and Astronomy, University of Birmingham, Edgbaston, Birmingham, B15 2TT, UK}
\date{Accepted 2022 April 8. Received 2022 April 7; in original form 2022 January 27}
\pubyear{2022}
\begin{document}
\label{firstpage}
\pagerange{\pageref{firstpage}--\pageref{lastpage}}
\maketitle
\begin{abstract}
We present the results of two-fluid hydrodynamical simulations of circumbinary discs consisting of gas and dust, with and without embedded planets, to examine the influence of the dust on the structure of the tidally truncated inner cavity and on the parking locations of migrating planets. In this proof-of-concept study, we consider Kepler-16 and -34 analogues, and examine dust fluids with Stokes numbers in the range $10^{-4} \le \Stokes \le 10^{-1}$ and \dtg ratios of 0.01 and 1. For the canonical \dtg ratio of 0.01, we find the inclusion of the dust has only a minor effect on the cavity and stopping locations of embedded planets compared to dust-free simulations. However, for the enhanced \dtg ratio of unity, assumed to arise because of significant dust drift and accumulation, we find that the dust can have a dramatic effect by shrinking and circularising the inner cavity, which brings the parking locations of planets closer to the central binary. This work demonstrates the importance of considering both gas and dust in studies of circumbinary discs and planets, and provides a potential means of explaining the orbital properties of circumbinary planets such as Kepler-34b, which have hitherto been difficult to explain using gas-only hydrodynamical simulations.
\end{abstract}

\section{Introduction} 
\label{sec:intro}

Circumbinary discs are accretion discs surrounding two central objects, e.g. two black holes or binary stars.
In recent years 13 circumbinary planets have been discovered orbiting binary stars by the Kepler and TESS spacecraft (e.g. Kepler-16b\footnote{Kepler-16b was recently detected using the radial velocity technique \citep{Triaud21}, and the derived parameters are consistent with those obtained from the Kepler data.} \citep{Doyle11}, Kepler-34b \citep{Welsh12}, TOI-1338b \citep{Kostov20}), raising many questions about their formation and the evolution of the circumbinary discs they are thought to form within \citep[see][for a review]{Martin18}.

Question about the formation and dynamics of circumbinary planets are long-standing.
Examining the locations where stable circumbinary orbits could exist, \citet{Dvorak89} derived a stability limit criterion based on numerical simulations of test particles around circumbinary stars.
Expanding on this work, \citet{Holman99} performed numerical simulations for much longer time-scales, determining that there exists a dynamical stability limit close to the central binary, where should objects orbit interior to this location, then gravitational perturbations from the binary stars would cause the orbit to go unstable, leading to ejection of the planet from the system.
This stability limit depends mainly on the eccentricity of the binary, with more eccentric binaries being more disruptive.
Other studies have examined the effects of mutual inclinations between the binary stars and any orbiting test particles \citep{Pilat-Lohinger03,Doolin11}, finding that inclination has little effect on the location of the stability limit within the binary mass ratio and eccentricity parameter space.
The majority of observed circumbinary planets orbit close to this limit, with the main exceptions being the outer planets in the Kepler-47 multi-planet system \citep{Orosz12_k47,Orosz19} and Kepler-1647b \citep{Kostov16}.
The propensity for these planets to orbit close to this location, as well as their inferred coplanarity with the binary's orbital plane, indicates a common formation pathway for these planets within coplanar circumbinary discs.

Two possible formation pathways are: {\it in situ} formation from material surrounding the binary; or formation at a larger distance and then migration to their observed orbits.
The {\it in situ} model suffers from numerous issues that hinder the formation of planets near the stability limit, including: differential pericentre alignment of eccentric planetesimals of different sizes leads to corrosive collisions \citep{Scholl07}; gravitational interactions with non-axisymmetric features within circumbinary discs leading to large impact velocities between planetesimals \citep{Marzari08,Kley10}; excitation of planetesimal eccentricities through N-body interactions lead to large relative velocities, that are disruptive for accretion onto planetary bodies \citep{Meschiari12a,Meschiari12b,Paardekooper12,Lines14,Bromley15}.
More recently it has been shown that {\it in situ} pebble accretion scenarios also suffer from difficulties because a parametric instability can generate hydrodynamical turbulence that stirs up pebbles, rendering pebble accretion inefficient \citep{Pierens20,Pierens21}. 
These issues, alongside the proximity of these planets to their host binary’s stability limit and their coplanarity, suggest they formed further out in the disc and moved to their current orbits through disc-driven migration.

Numerous works have shown that migrating planets in circumbinary discs stall when they reach the central cavity. The precise stopping location depends on parameters such as the planet mass. Giant planets open a gap in the disc and circularise the inner cavity, and tend to park closer to the binary \citep[(although this increases the probability they may be ejected)][]{Nelson03,Pierens08a,Thun18}.
Lower mass planets migrate to the inner cavity edge and their migration ceases due to a strong corotation torque counteracting the Lindblad torques \citep{Pierens07,Pierens08b}.
The orbits of these low mass planets align with the inner eccentric disc and precess with it in a state of apsidal corotation \citep{Thun18}.
Other works have found that the stopping locations and planet eccentricities are influenced by the mass of the discs that they form in \citep{Dunhill13} or by the effects of gas self-gravity on the disc structure \citep{Mutter17P}.
More recently, \citet{Penzlin21} found that disc parameters including the viscosity parameter $\alpha$ and the disc aspect ratio $H/R$ also affect the stopping locations of migrating planets, mainly by permitting the planets to open partial gaps more easily in the disc, forcing the disc to become more circularised, allowing the planets to migrate closer to central binary. In attempting to match the locations of known circumbinary planets using hydrodynamical simulations, a long standing problem has been to match systems such as Kepler-34b because the central cavity that forms is very large and eccentric in this case, causing the planet to park too far away from the central binary \citep[e.g.][]{Pierens13, Penzlin21}.

Other works have examined the evolution and structure of circumbinary discs without embedded planets.
It has long been accepted that tidal torques exerted from a central binary on to its circumbinary disc creates a central cavity.
The radial extent of this cavity depends on a number of factors, including the binary mass ratio, eccentricity and disc parameters \citep{Artymowicz94}.
Direct imaging of the binary GG Tau system shows an inner cavity to its circumbinary disc agreeing with theoretical studies \citep{Dutrey94, Keppler2020}.
Further studies have found that interactions between the disc and the binary produces an asymmetric, eccentric precessing disc \citep{Pelupessy13,Pierens13,Kley14,Thun17,Mutter17D}.
More recently, \citet{Thun17} found that cavity sizes and eccentricities only change as function of the binary eccentricity as well as the local disc properties, while the precession rate also varies with respect to the mass of the binary.
Following on from the work of \citet{Thun17}, \citet{Kley19} examined the properties of circumbinary discs with a more realistic treatment of the thermal structure, by including viscous heating and radiative cooling, finding good agreement with locally isothermal simulations for inner disc regions and cavity sizes and eccentricities.

Recently, \citet{Chachan19} used smoothed particle hydrodynamic simulations to explore the evolution of dust in circumbinary discs, finding that large uncoupled dust grains become trapped at the edge of the cavity.
With large amounts of dust becoming trapped there, this could possibly favour planet formation through {\it in situ} accretion.
A caveat of these simulations, is that they were performed in the test-particle limit, neglecting the back-reaction forces of the dust on the gas.
Including the back reaction of dust on to the gas can have important consequences on the disc structure and evolution, especially when the local dust-to-gas ratios approach unity.

In this paper, we investigate the impact that dust has on the evolution and structure of circumbinary discs, as well as on the parking locations of migrating planets. In this first proof-of-concept study, we use the default setting in the {\sc{fargo3D}} multi-fluid code and include dust species with fixed Stokes numbers. In each simulation we assume that an initial global abundance of dust is present, defined by the \dtg ratio, in the form of a single dust species with a fixed Stokes number. Our work is motivated by the fact that observations show that dust growth and drift is occurring in protoplanetary discs, such that there should be an accumulation of dust near the inner cavity in a circumbinary disc, and some systems such as Kepler-34b have been difficult to match using gas-only hydrodynamical simulations. Our work investigates whether or not the back-reaction of the dust on the gas influences the structure of the cavity and parking locations of planets, and we find in some cases that the dust has a dramatic effect. This result motivates further studies using more realistic set-ups that will be presented in future publications.

This paper is organised as follows.
Section \ref{sec:method} outlines the physical model and initial conditions used in the simulations.
Section \ref{sec:disc_evol} examines the effects that the inclusion of dust has on the structure and evolution of circumbinary discs.
The evolution of embedded planets is then presented in sect. \ref{sec:planet_evol} before we draw our conclusions in sect. \ref{sec:conclusions}.

\section{Method}
\label{sec:method}
\subsection{Numerical Method}
We run 2D multi-fluid hydrodynamical simulations where we evolve both the gas and dust simultaneously using {\sc{fargo3D}} \citep{FARGO-3D-2016, FARGO-3D-2019}.
We detail the equations of motion below using 2D polar coordinates ($R, \phi$) centred on the centre of mass of the binary system.
For the 2D approximation we consider vertically integrated quantities, such as the surface density $\Sigma = \int^{\infty}_{-\infty} \rho~ \mathrm{d}z$.
With this set-up, the continuity equation is given by
\begin{equation}
    \dfrac{\partial \Sigma_i}{\partial t} + \nabla\,\mathbf{\cdot}\,(\Sigma_i \mathbf{v}_i) = 0.
\end{equation}
The momentum equations in the radial and azimuthal directions are
\begin{equation}
	\dfrac{\partial \Sigma_i \upsilon_{R,i}}{\partial t} + \nabla\,\mathbf{\cdot}\,(\Sigma_i \upsilon_{R,i}\mathbf{v}_i) - \frac{\Sigma_i \upsilon_{\phi,i}^2}{R} = -\dfrac{\partial p_i}{\partial R} - \Sigma \dfrac{\partial \Phi}{\partial R} + f_R + \mathbf{F}_{R,i}
\label{eq:v_rad}
\end{equation}
and
\begin{align}
\begin{split}
	\dfrac{\partial \Sigma_i \upsilon_{\phi,i}}{\partial t} + \nabla\,\mathbf{\cdot}\,(\Sigma_i \upsilon_{\phi,i}\mathbf{v}_i) + \frac{\Sigma_i \upsilon_{R,i} \upsilon_{\phi,i}}{R} \\
	= -\frac{1}{R}\dfrac{\partial p_i}{\partial \phi} - \frac{\Sigma}{R} \dfrac{\partial \Phi}{\partial \phi} + f_{\phi} + \mathbf{F}_{\phi,i}.
	\end{split}
\end{align}
The subscript $i$ is the index of the species, i.e. dust or gas, $p_i$ is the vertically averaged pressure, which is zero for the pressure-less dust component, ($v_{\rm R},v_{\phi}$) are the radial and azimuthal components of the velocity, $\mathbf{v}_i$, $f_{R,i}$ and $f_{\phi,i}$ are the components of the vertically-averaged viscous force per unit volume (described in \citet{Nelson2000}) that are non-zero only for the gas component.
The changing gravitational potential experienced by the disc arising from the binary system and embedded planets is accounted for by $\Phi$.
We do not include the gravitational back-reaction nor accretion from the disc on to the binary, so the binary parameters remain fixed.
The drag force per unit volume $\mathbf{F}_i$ is defined as
\begin{equation}
    \mathbf{F}_i = -\rho_i \sum_{j\neq i}\alpha_{ij}(\mathbf{v}_i-\mathbf{v}_j),
\end{equation}
with $\alpha_{ij}$ being the collision rate between species $i$ and $j$.
This collision rate parameterises the momentum transfer per unit time and is, in general, a function of the physical properties of the dust species, i.e. size, and the relative gas-dust velocity.
Referring to the gas species with the index $g$, the collision rate is written as
\begin{equation}
    \alpha_{i,j} =\alpha_i\delta_{jg} + \epsilon_j\alpha_j\delta_{ig},
\end{equation}
with $\epsilon_j =\rho_j/\rho_g$ and $\delta_{ig}$ being the Kronecker delta.
The collision rate is parameterised by the Stokes number, $St$, a dimensionless parameter that characterises the collision rate in units of the local Keplerian frequency $\Omega_{\rm K}$
\begin{equation}
    \alpha_i = \frac{\Omega_{\rm K}}{St_i}.
\end{equation}
The Stokes number depends on the properties of the gas, the dust grains, and their relative velocity \citep{Safronov72,Whipple72}.
For the simulations in this paper, which are intended to provide an initial exploration of how the concentration of dust at the inner cavity edge of the disc affects its dynamical evolution, we assume that the Stokes number remains constant in space and time. The effects of implementing a more sophisticated treatment of the dust will be explored in forthcoming work.

\subsection{Boundary conditions}
\label{sec:BCs}
Ideally, the computational domain would incorporate each member of the central binary. However, this is not computationally feasible because of the very small timesteps needed when calculating the evolution of gas close to either star. Therefore, to make the simulations tractable, it is necessary to choose the inner boundary location such that it encloses one or both members of the binary, whilst simultaneously ensuring that the simulation results are not significantly affected. Recent studies \citep{Mutter17D,Thun17} have highlighted the strong dependence of the structure of circumbinary discs on the choice of the inner boundary condition and on its location.
Choosing between open or closed boundary conditions, or a boundary condition that allows outflow at a rate corresponding to the viscous flow speed, can lead to different cavity sizes and eccentricities \citep{Mutter17D}. Comparing simulations where the binary is incorporated in the computational domain with runs where the binary sits within the boundary shows that consistent results are obtained when the location of the inner boundary is comparable to the binary semi-major axis \citep{Thun17}.

Taking these dependencies into account, we use an open inner boundary condition and place it at a distance equal to the binary semi-major axis.
This allows gas to flow out of the disc and prevents inflow. We use a closed outer boundary condition and employ a wave killing zone to avoid wave reflection.

\subsection{Simulation set-up}
\label{sec:setup}
The discs in our simulations extend out to 20 $\au$.
The gas surface density is initialised according to
\begin{equation}
    \Sigma_{\rm g}(R) = f_{\rm gap}\Sigma_0 \left(\frac{R}{5.2\au}\right)^{-1}
\end{equation}
and the dust surface density is given by
\begin{equation}
    \Sigma_{\rm d}(R) = \epsilon \Sigma_{\rm g},
\end{equation}
where $\epsilon$ is the initial \dtg ratio, and $f_{\rm gap}$ is a function that we use to establish an inner cavity in order to decrease the time taken for the discs to reach equilibrium.
We follow \citet{Gunther04} in calculating $f_{\rm gap}$
\begin{equation}
    f_{\rm gap} = \left(1 + \exp{\left[-\frac{R-R_{\rm gap}}{0.1 R_{\rm gap}}\right]}\right)^{-1},
\end{equation}
where $R_{\rm gap} = 2.5 \ab$, which is the expected gap size created by the tidal interaction of the binary on the disc \citep{Artymowicz94}.

We set $\Sigma_0$ such that the gas disc mass within 40 $\au$ is equal to 5$\%$ of the combined binary mass. For the viscosity we employ the alpha prescription \citep{Shak} $\nu = \alpha c_{\rm s} H$, with $\alpha=10^{-3}$, where $c_{\rm s}$ is the sound speed and $H$ is the local disc thickness.
We consider discs with a constant disc aspect ratio $h=0.05$, and we adopt a locally isothermal equation of state, with a temperature profile given by $T=T_0 R^{-1}$.
This choice of equation of state allows the results presented here to be compared with other similar studies \citep[e.g.][]{Pierens13,Mutter17D,Mutter17P} such that the effects of including the dust and its back reaction on the gas can be examined.

For the numerical grid, we use a resolution of $N_{\rm r} \times N_{\rm \phi} = 768 \times 512$, with logarithmic radial spacing.
For stability and computational reasons, especially in the inner cavity where the density can drop to very low values, we impose a density floor equal to $\Sigma_{\rm floor} = 10^{-4}$gcm$^{-2}$.
This floor is applied to both the gas and the dust surface densities and is generally at least four orders of magnitude smaller than the densities outside of the inner cavity region.

To explore the effects that dust has on the disc structure, we assign the dust a different Stokes numbers, $\Stokes$, for each simulation.
We use Stokes numbers ranging from $\Stokes=10^{-4}$, corresponding to small dust that is strongly coupled to the gas, to $\Stokes=10^{-1}$, corresponding to larger dust that is less strongly coupled to the gas and which results in faster radial drift of the dust. For reference, the particle size corresponding to the Stokes number for a gas surface density of 1000~g~cm$^{-2}$ (typical near the cavity edge) is given by $a \sim 200 \Stokes$~cm \citep{Birnstiel2010}.
We also change the initial \dtg ratio $\epsilon$ with values being equal to either 0.01 or 1.
This allows us to compare the results for discs where there is a high concentration of dust in the central regions to discs where the central regions are relatively dust poor. Note the choice of $\epsilon=1$ is not meant to represent discs with a global \dtg value of unity, but instead provides a means of examining what happens if the inner regions of discs become dust rich because of large scale radial drift and accumulation in the inner regions without having to run the simulations for unfeasibly long times.

With the larger values of the Stokes numbers, it is important to estimate the radial drift timescales from the outer regions of the disc. With our choice of setup, dust with Stokes numbers of 0.1 has a drift timescale of $\tau_{\rm d}>3\times 10^{5}$ binary orbits at $r=60\au$, and $\tau_{\rm d}>5\times 10^{5}$ binary orbits at the disc outer edge. These long timescales allow the inner disc to be continually fed from the outer disc throughout the simulation. For dust with smaller Stokes numbers, the timescales are correspondingly longer.

In sect. \ref{sec:planet_evol} we include planets within the simulations to examine the effects that the inclusion of dust has on their migration and stopping locations.
When considering the drift timescales at the injection locations of the planets, we find that for the largest dust species and dust-to-gas ratio, they are similar to planet migration timescales, whilst for smaller dust sizes and gas-to-dust ratios, migration timescales are always shorter than the radial drift timescales.
A full list of our binary and simulation parameters can be found in table \ref{tab:systems}.

\begin{table}
	\centering
	\caption{Binary, planet and disc parameters.}
	\begin{tabular}{ccc}
		\hline
		\hline
		 & Kepler-16 & Kepler-34\\
		\hline
		$M_\mathrm{A}\ (M_\odot)$ & 0.690 & 1.048\\
		$M_\mathrm{B}\ (M_\odot)$ & 0.203 & 1.021\\
		$a_\mathrm{b}$ (au) & 0.224 & 0.228\\
		$e_{\rm b}$ &	 0.159 & 0.521 \\
		$a_\mathrm{p}$ (au) & 0.705 & 1.090\\
		$e_{\rm p}$ &	 0.007 & 0.182\\
		Planet release time ($P_{b}$) & 30,000 & 50,000 \\
    	$M_{\rm p, init} (\me)$ & \multicolumn{2}{c}{20} \\
		N$_{\rm r}$  & \multicolumn{2}{c}{768} \\
        N$_{\rm \phi}$  & \multicolumn{2}{c}{512} \\
        R$_{\rm in}$ & \multicolumn{2}{c}{$1\times \ab$} \\
        R$_{\rm out} (\au)$ & \multicolumn{2}{c}{20} \\
        Stokes Values ($St$) & \multicolumn{2}{c}{[$10^{-4}\, 10^{-3}\, 10^{-2}\, 10^{-1}$]} \\
        Dust-to-gas Ratio $\epsilon$ & \multicolumn{2}{c}{0.01, 1} \\
        disc aspect ratio $h$ & \multicolumn{2}{c}{0.05} \\
        viscosity parameter $\alpha$ & \multicolumn{2}{c}{$10^{-3}$} \\
        $\Sigma_0(5.2\au)$ (gcm$^{-2}$)& 305.13 & 707.531 \\ 
        Reference &\citet{Doyle11}&\citet{Welsh12}\\
		\hline
	\end{tabular}
	\label{tab:systems}
\end{table}

\section{Disc Evolution}
\label{sec:disc_evol}
Using the disc set-up discussed in the previous section we ran a set of simulations examining the role the inclusion of dust of different sizes and abundances has on the structure and dynamics of the circumbinary discs in Kepler-16 and -34 analogues.
As the size and abundance of the dust increases the influence on the gas increases, hence altering the structure of the disc when compared to a dust-free simulation.
Using Stokes numbers between $10^{-4}$ and $10^{-1}$ and \dtg ratios of 0.01 and 1, we examine the effects of both loosely coupled and tightly coupled dust, as well as dust-rich discs and those with standard dust abundances.

\subsection{Calculating cavity and disc properties}
\label{sec:calc_properties}

Interactions between binary stars and their circumbinary discs lead to the formation of a tidally-truncated eccentric inner cavity, extending out numerous binary separations past the stability limit \citep{Holman99}.
To determine the effects that dust has on this inner cavity, we first need to calculate the properties of the gas in this region and the surrounding disc.

To calculate the cavity size (semi-major axis) and eccentricity, we follow \citet{Thun17} where it is assumed that the focus of an ellipse imposed on the cavity is located at the centre of mass of the binary.
With this assumption, we then determine the coordinates ($R_{\Sigma \rm max},\phi_{\Sigma \rm max}$) of the cell with the largest surface density value, which defines the direction to the cavity apocentre.
We then define the apastron of the cavity $R_{\rm apo}$ as the minimum radius along the line ($R,\phi_{\Sigma \rm max}$) which meets the condition
\begin{equation}
    \Sigma(R,\phi_{\Sigma \rm max}) \ge 0.1 \times \Sigma(R_{\Sigma \rm max},\phi_{\Sigma \rm max}).
\end{equation}
The cavity periastron, $R_{\rm per}$, is then defined as the minimum radius along the line ($R,\phi_{\Sigma \rm max}+\pi$) in the opposite direction that fulfils
\begin{equation}
    \Sigma(R,\phi_{\Sigma \rm max}+\pi) \ge 0.1 \times \Sigma(R_{\Sigma \rm max},\phi_{\Sigma \rm max}).
\end{equation}
With the periastron and apastron now calculated, we can find the semi-major axis and eccentricity of the cavity through
\begin{equation}
\label{eq:cav_size}
    a_{\rm cav} = 0.5(R_{\rm per}+R_{\rm apo}),
\end{equation}
\begin{equation}
\label{eq:cav_ecc}
    e_{\rm cav} = R_{\rm apo}/a_{\rm cav}-1.
\end{equation}

However, since the cavity eccentricity only gives the eccentricity of what we define as the cavity edge, it does not give the eccentricity profile throughout the disc, which could also highlight effects from the dust acting on the gas.
To calculate the disc eccentricity radial profile, we first calculate the orbital elements of each grid cell by treating them as particles, with the mass and velocity of the cell orbiting the binary.
We then follow \citet{Pierens13} to calculate the azimuthally averaged eccentricity:
\begin{equation}
\label{eq:disc_ecc_r}
    e(R) = \dfrac{\int^{2\pi}_{0}\int^{R+dR}_{R}\Sigma~e_{\rm c}~R~{\rm d}R~{\rm d}\phi}{\int^{2\pi}_{0}\int^{R+dR}_{R}\Sigma~R~{\rm d}R~{\rm d}\phi},
\end{equation}
where $\Sigma$ is the grid cell surface density, $e_{\rm c}$ is the calculated eccentricity of the cell as a particle.
We then also calculate the eccentricity of the inner disc region,
\begin{equation}
\label{eq:disc_ecc_d}
    e_{\rm d} = \dfrac{\int^{2\pi}_{0}\int^{R_2}_{R_1}\Sigma~e_{\rm c}~R~{\rm d}R~{\rm d}\phi}{\int^{2\pi}_{0}\int^{R_2}_{R_1}\Sigma~R~{\rm d}R~{\rm d}\phi},
\end{equation}
where $R_1$ is the inner disc radius, and $R_2$ is the outer disc radius that we calculate the disc eccentricity for.
Since the disc extends out much further than the cavity region here, a significant fraction of the outer disc will be unperturbed by the binary and as such will be on essentially circular orbits.
Including such locations in the calculation of $e_{\rm d}$ will artificially reduce its magnitude, limiting it as a guide of the disc eccentricity around the cavity region.
Therefore we set an arbitrary value for $R_2$ for both Kepler-16 and Kepler-34 where the perturbations from the binary are seen to have minimal effects on the local disc eccentricities.
For Kepler-16 we take $R_2=20\ab$, and later for Kepler-34 we take $R_2=30\ab$ since the Kepler-34 binary influences a larger region of the disc.

Below we now present the results for each binary system in turn, where we highlight the differences that arise as a result of the changing dust properties.

\subsection{Kepler-16}
\label{sec:kep-16_disc}

\subsubsection{Cavity properties}
\begin{figure}
    \centering
    \includegraphics[scale=0.4]{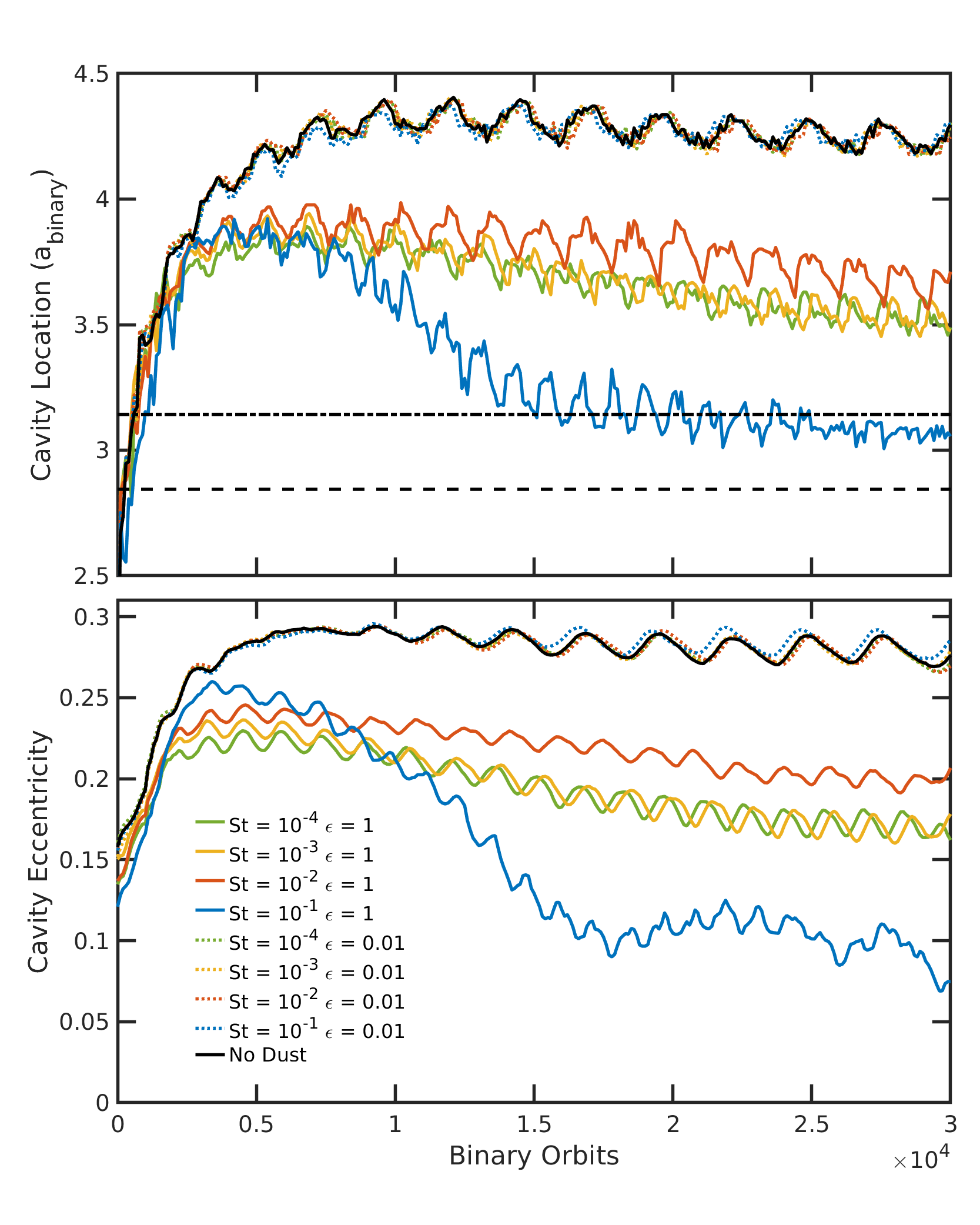}
    \caption{Time evolution of the cavity location (top panel) and eccentricity (bottom panel) for the discs in the Kepler-16 system. Coloured lines represent different Stokes numbers for the dust: blue being Stokes number of 0.1, red being $10^{-2}$, yellow being $10^{-3}$ and green being $10^{-4}$. The Solid black shows the disc where dust was not included. We differentiate between the two \dtg ratios by using solid coloured lines for \dtg ratios of 1, and dotted lines for \dtg ratios of 0.01. The horizontal dashed and dot-dashed line represent the stability limit \citep{Holman99} and the observed location of Kepler-16b \citep{Doyle11} respectively.}
    \label{fig:kep-16_cav_sma_ecc}
\end{figure}

The top panel of fig. \ref{fig:kep-16_cav_sma_ecc} shows the evolving cavity sizes for the discs containing dust of different Stokes numbers and abundances, where each model is described in the figure caption and legend.
The black dashed and dot-dashed horizontal lines denote the locations of the stability limit \citep{Holman99} and of Kepler-16b \citep{Doyle11} respectively.

For all of the models in the top panel of fig. \ref{fig:kep-16_cav_sma_ecc}, it takes $\sim 10,000$ binary orbits for the cavity to reach a steady-state.
For the models where the \dtg ratio is equal to 0.01, they all lie beneath the black line that represents the dust-free disc that has a cavity size of $\sim 4.3\ab$, significantly further away from the central binary than the stability limit as well as the observed location of Kepler-16b.
Whilst the cavity semi-major axis may be located far away from the stability limit, the cavity pericentres for these discs are all close to the location of the stability limit, making it a good proxy for the cavity pericentre. Gas on orbits interior to the limit is associated with streamer channels along which gas accretes onto the binary stars.
For the discs with low \dtg ratios, the cavity sizes are independent of the Stokes number of the dust, showing that for these discs the low dust abundance has only a minimal effect on the dynamics of the gas around the cavity edge.

Increasing the dust abundance so the \dtg ratio equals 1 causes the cavities to shrink compared to the dust-free discs, as seen when comparing the solid coloured lines to the black line in the top panel of fig. \ref{fig:kep-16_cav_sma_ecc}.
With the dust being more abundant in these discs, the back reaction from the dust acts to remove orbital energy from the gas and circularises it, whilst the pericentres remain roughly fixed around the stability limit.
The effect of having different Stokes numbers, also becomes apparent here, where the back-reaction from the different dust sizes onto the gas now plays a more significant role.
All of these dusty discs again take $\sim 10,000$ binary orbits to reach an equilibrium state with a cavity size of $\sim 3.8\ab$.
Then, over the next 20,000 binary orbits, the cavity sizes begin to decrease, with the more coupled dust ($\Stokes$ of $10^{-4}$ to $10^{-2}$) achieving cavity sizes of around $3.5\ab$.
The most weakly coupled dust ($\Stokes~0.1$) has the strongest influence on the gas and forces the cavity size to drop to around $3\ab$, just exterior to the stability limit.
With the pericentre of the gas still located near the stability limit, this now being close to the cavity size, the cavity becomes more circular. This longer term evolution arises because of the inwards drift of the dust from further out in the disc, increasing the \dtg ratio near the cavity.

This circularisation of the dust-rich discs can be seen in the bottom panel of fig. \ref{fig:kep-16_cav_sma_ecc} where we plot the eccentricity of the cavity as the disc evolves.
For the standard dust abundance (\dtg ratios of 0.01), all of the dotted lines sit more or less beneath the black line, with the cavities reaching eccentricities between 0.25 and 0.3.
The effects of increasing the dust abundance on the cavity eccentricities are apparent, with all of the solid lines only reaching a maximum eccentricity of $\sim0.25$.
Whilst the discs with more coupled dust settle into a steady solution, with final eccentricities of 0.16--0.2, the cavity eccentricity in the disc with dust of $\Stokes = 0.1$ quickly drops to around 0.1 and then after 30,000 binary orbits to around 0.07, much more circular than its counterparts. As mentioned above, this time evolution is caused by the continuous drift and accumulation of the larger dust at the cavity edge.

We note that when the \dtg ratio equals 1 the degree of shrinkage and circularisation of the cavities is not strictly monotonic with changes in the Stokes number. Figure~\ref{fig:kep-16_cav_sma_ecc} shows that the cavity size and eccentricity increases slightly when the Stokes number increases from $10^{-3}$ to $10^{-2}$, before decreasing dramatically for $\Stokes=0.1$. We have investigated this further by running additional simulations for intermediate values of the Stokes number. We observe that changes in the cavity structure as a function of the Stokes number are continuous, indicating the non-monotonic behaviour is a real effect, but at present we do not have a physical explanation for it. We will explore this effect further in future work.

\begin{figure}
    \centering
    \includegraphics[scale=0.6]{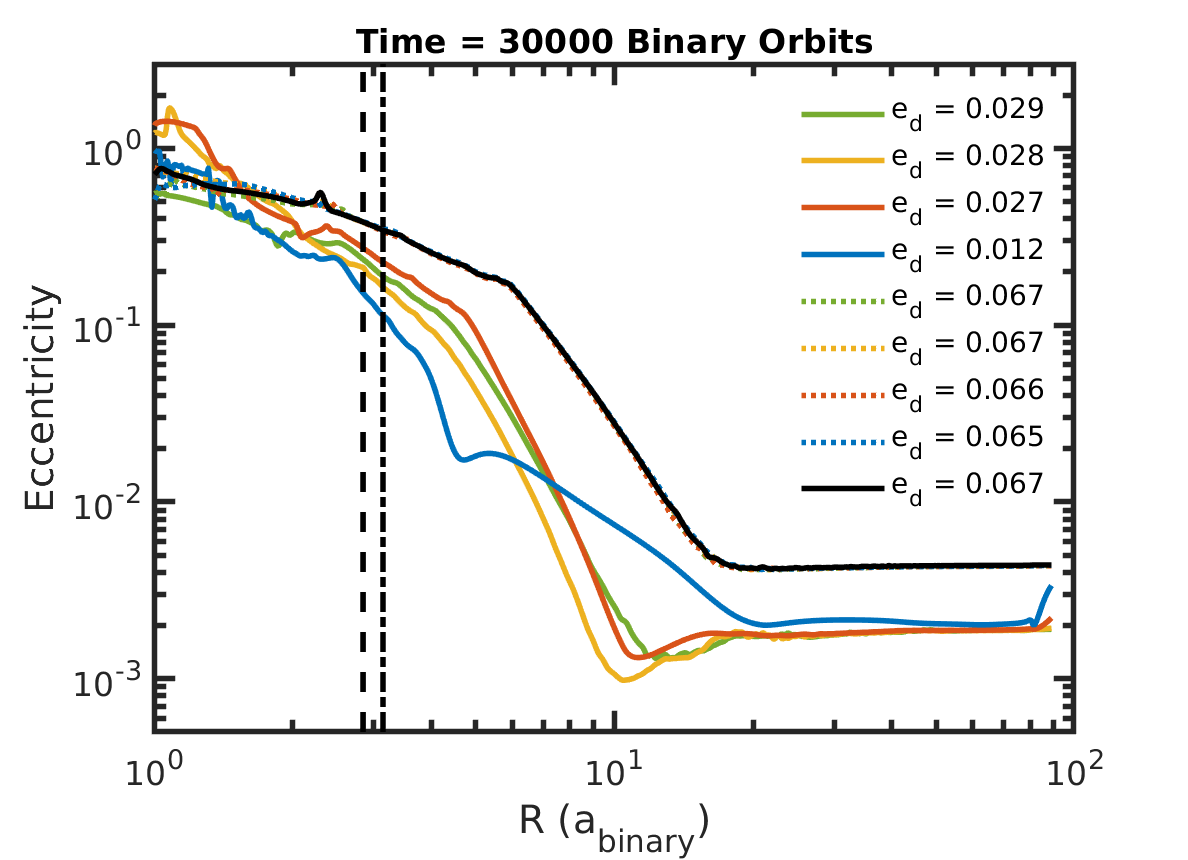}
    \caption{Eccentricity profiles as a function of orbital radius for the discs in the Kepler-16 system. Colours and line formats are the same as in fig. \ref{fig:kep-16_cav_sma_ecc}. The values in the legend denote the time-averaged disc eccentricity up to an orbital distance of 20$\ab$.}
    \label{fig:kep-16_disc_ecc}
\end{figure}

\begin{figure*}
    \centering
    \includegraphics[scale=0.45]{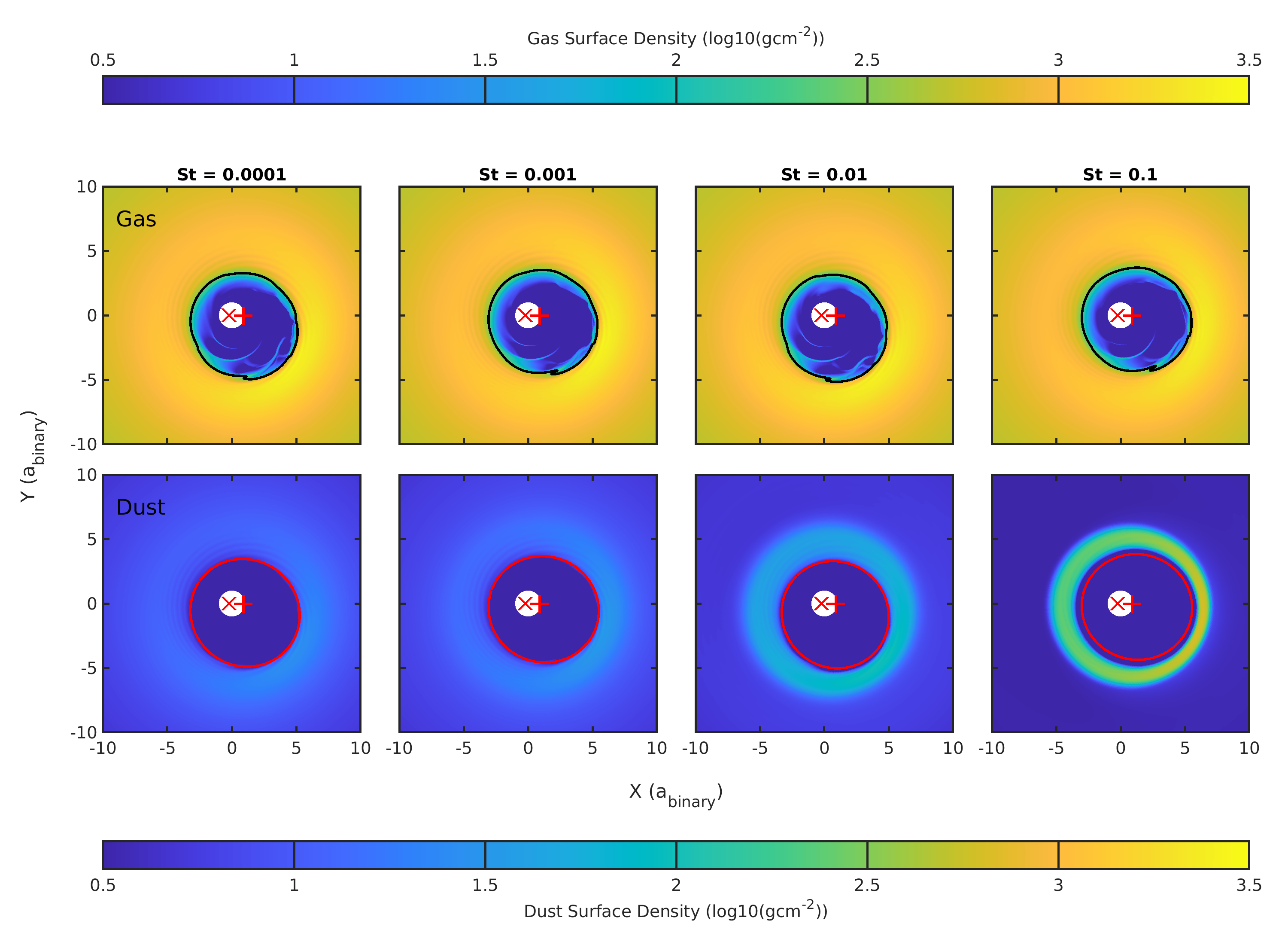}
    \caption{Two-dimensional surface density plots of the gas (top panels) and dust (bottom panels) for the discs in the Kepler-16 system.
    The \dtg ratio for the discs was equal to 0.01. Going from left-to-right we increase the Stokes number of the dust from $10^{-4}$ to $10^{-1}$, shown at the top of each column. The red `x' and `+' denote the locations of the binary stars, with the black line in the top panels showing the rough location of the cavity with a contour of the surface density at a value equal to $10\%$ of the maximum surface density. The red ellipse in the bottom panels shows the gas cavities using eq. \ref{eq:cav_size}.}
    \label{fig:kep-16_2d_0.01}
\end{figure*}

Figure \ref{fig:kep-16_disc_ecc} shows the azimuthally averaged eccentricities using equation \ref{eq:disc_ecc_r} at a time of 30,000 binary orbits.
The colour coding and line formats are the same as used in fig. \ref{fig:kep-16_cav_sma_ecc}, with the legend showing a time-averaged disc eccentricity (eq. \ref{eq:disc_ecc_d}) over the previous 5,000 binary orbits for each disc.
Generally, all of the discs have qualitatively similar profiles, highlighting the dominant effect of the perturbations from the binary.
Looking at the black line for the dust-free disc, the eccentricity profile can be split into three regions: inside the cavity; just exterior to the cavity; and far from the binary.
Interior to the cavity, the eccentricity is understandably high, since the perturbations from the binary stars are strongest there.
Moving further out, the perturbations from the binary decrease, causing a reduction in the eccentricity.
This continues at a consistent rate up to an orbital radius just exterior to where the cavity apocentre resides.
For the black line, this is at an orbital radius of $\sim6\ab$, and can be seen as a `knee' in the eccentricity profile.
Exterior to the cavity region, the eccentricity drops more steeply, with the binary perturbations reducing further in magnitude.
This continues until the eccentricity induced by the binary perturbations becomes weaker than the gas pressure support, occurring at around $15\ab$ for the black line, where exterior to this location the finite eccentricity measurement originates from the influence of the pressure support on the equilibrium gas velocities and does not indicate that the fluid streamlines are non-circular.
We note the azimuthal velocities for the disc containing no dust in the region of dominant pressure support are 0.2$\%$ slower than the Keplerian velocity, i.e. sub-Keplerian, but still appear as circular orbits. This value is also found for the other discs with low dust-to-gas ratios.

As with the calculations for the cavity sizes and eccentricities, the discs with low \dtg ratios display similar eccentricity profiles to the disc with no dust.
In fig. \ref{fig:kep-16_disc_ecc}, all of the dotted lines showing the discs with low \dtg ratios lie beneath the black line.
When looking at the time-averaged disc eccentricities for the inner disc region, interior to $20\ab$ for Kepler-16, the legend shows negligible differences between the calculated values, ranging from 0.065--0.067 for the discs with either no dust or standard dust abundances.

\begin{figure*}
    \centering
    \includegraphics[scale=0.45]{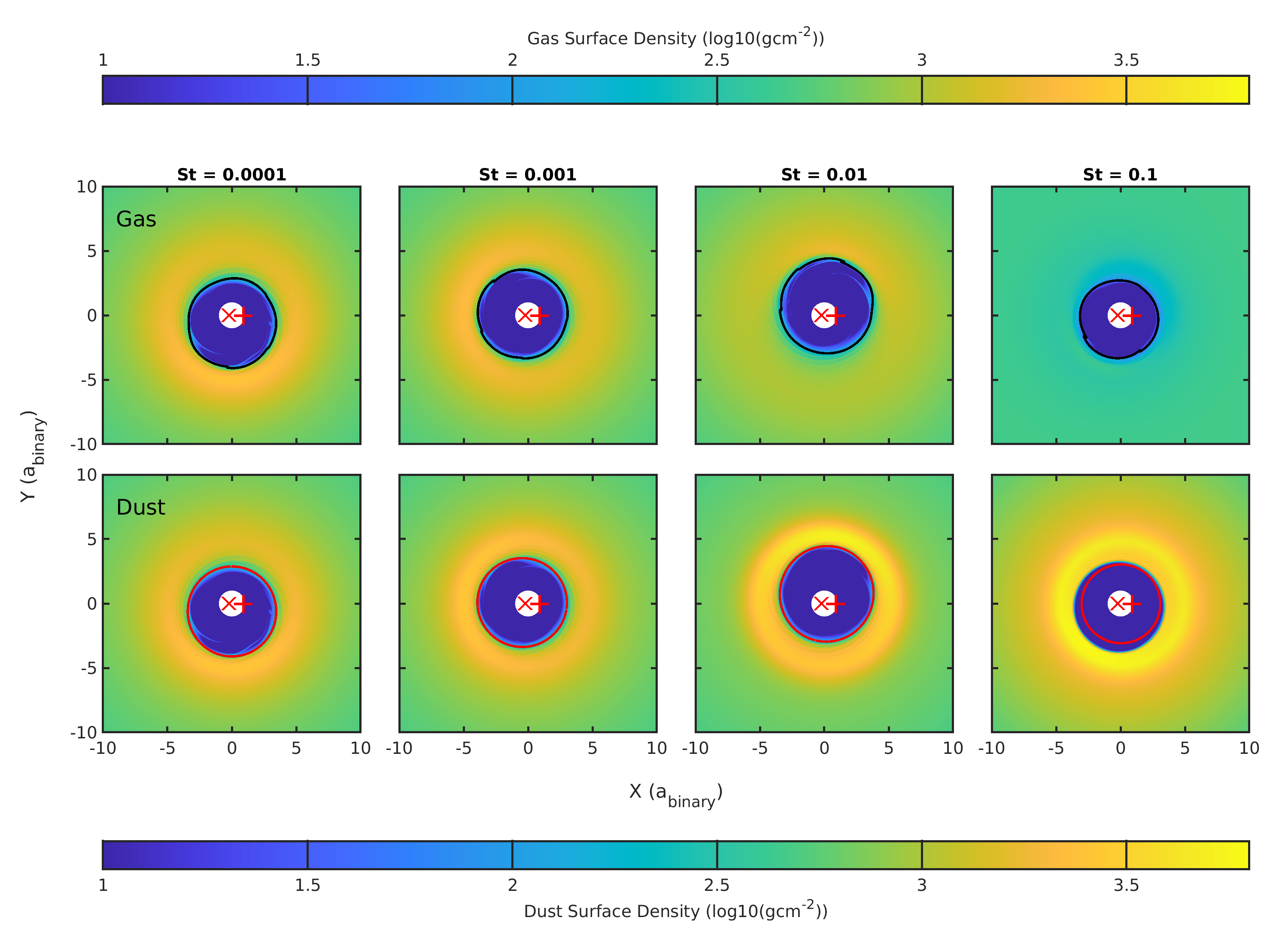}
    \caption{Same as fig. \ref{fig:kep-16_2d_0.01} but for discs with \dtg ratios equal to 1.}
    \label{fig:kep-16_2d_1}
\end{figure*}

Increasing the \dtg ratio from 0.01 to 1, the solid coloured lines in fig. \ref{fig:kep-16_disc_ecc} show that the discs are less eccentric, as expected from the cavity eccentricities plotted in fig. \ref{fig:kep-16_cav_sma_ecc}.
This occurs throughout the disc, with the back-reaction of the dust acting to circularise the gas orbits.
Whilst the qualitative behaviour discussed above remains similar, with high eccentricities due to binary perturbations inside the cavity, and then dropping off exterior to the cavity apocentre until they achieve a small finite value because of pressure support in the gas, the eccentricities are quantitatively smaller.
This can also be seen in the legend of fig. \ref{fig:kep-16_disc_ecc}, where the time averaged disc eccentricities for the solid coloured lines ($e_{\rm d}<0.03$) are much smaller than the black line representing the disc with no dust ($e_{\rm d}=0.067$).
The effect of having different sized dust also becomes more evident for the coloured solid lines in fig. \ref{fig:kep-16_disc_ecc}, with the disc containing dust with $\Stokes=0.1$ showing numerous differences in the eccentricity profile.
With the cavity becoming circular, the knee in the eccentricity profile has disappeared with there now being a dip around that location ($\sim4$--$5\ab$).
This dip is situated around the cavity apocentre, at the location where the dust is most abundant, acting to circularise the cavity.
The circularisation is also seen in the time-averaged eccentricity for the inner disc region where it averages out at 0.012, considerably smaller than the disc with no dust (0.067) as well as the other discs with unity \dtg ratios (0.027--0.029).
When examining the outer discs, where they are dominated by pressure support, the eccentricities are slightly lower than in the discs with low dust-to-gas ratios, since the back reaction from the dust on the gas causes the gas to orbit slightly closer to the Keplerian velocity, resulting in smaller values of the measured eccentricities.
Even though the eccentricities are dominated by the pressure support, the dust size has a more minor effect by further slowing down the gas compared to Keplerian, with larger dust sizes inducing larger effects.
Indeed the reductions in velocities due to the dust size is generally $<1\%$ of those arising from pressure support.

In summary, figs. \ref{fig:kep-16_cav_sma_ecc}--\ref{fig:kep-16_disc_ecc} show that as the Stokes numbers or \dtg ratios increase, the gas cavity locations, their eccentricities, and the inner disc eccentricities all decrease as the dust removes energy and angular momentum from the gas. This effect is probably arising because, due to pressure forces, a dust-free gas disc is able to establish a well-defined precessing, eccentric mode near the cavity region that is forced by the binary. Pressure-less dust particles, however, cannot form such a mode and instead achieve moderate eccentricities due to secular forcing by the binary. The interaction between the dust and gas can then force the gas component to become increasingly circularised as the dust abundance increases, or as the Stokes number increases because this allows the dust to act more independently of the gas.

\subsubsection{Surface densities}
In addition to examining the orbital elements of the gas near the cavity, it is also useful to examine the 2D surface density profiles of the discs, as these can also display effects due to changing  the \dtg ratio and the Stokes number.
Figure \ref{fig:kep-16_2d_0.01} shows 2D surface density plots for the gas (top panels) and the dust (bottom panels) for the discs with \dtg ratios of 0.01 after 30,000 binary orbits.
The Stokes number, shown above the topmost panels, increases from left-to-right, with the colour schemes showing the logarithm of the gas (top bar) and dust (bottom bar) surface densities.
The binary stars are denoted by the red `+' and `x' at the centres of the panels.
In the top panels showing the gas, the solid black line shows a rough representation of the cavity using a contour equal to 10$\%$ of the maximum surface density.
The red line in the bottom panels showing the dust surface densities, shows the cavity size and orientation following eqs. \ref{eq:cav_size} and \ref{eq:cav_ecc}.
As can be seen, when comparing the red ellipses to the black contours, as well as the visual location of the cavity in the 2D plots, eqs. \ref{eq:cav_size} and \ref{eq:cav_ecc} give very good approximations to the cavity semi-major axes and eccentricities.

Looking at the gas surface densities in the top panels of fig. \ref{fig:kep-16_2d_0.01}, it is clear that the inclusion of dust with different Stokes numbers has little effect on the two-dimensional structures of the discs.
This is unsurprising given that there was very little difference in the cavity semi-major axes and eccentricities, as well as the radial eccentricity profiles, for these discs as shown by the dotted lines in figs. \ref{fig:kep-16_cav_sma_ecc}--\ref{fig:kep-16_disc_ecc}.
For these discs the dust is insufficiently abundant to significantly alter the gas profiles.

Whilst there is little change in the gas surface densities as a function of the Stokes number, increasing the Stokes number has very noticeable effects on the dust surface densities as shown in the bottom panels of fig. \ref{fig:kep-16_2d_0.01}.
In the bottom two left panels, for the strongly coupled dust of $\Stokes=10^{-4}$ and $10^{-3}$, the dust surface densities are roughly identical to the gas surface density, only two orders of magnitude smaller as expected for a \dtg ratio of 0.01.
The dust in these cases undergoes very slow radial drift and so there are minimal build-ups around the cavity, with the dust interior to the cavity following the gas streamers and accreting onto the binary stars.
Moving to the panel with $\Stokes=10^{-2}$, there is a clear increase in the dust surface density around the cavity edge due to the increased drift velocity of the larger dust grains. 
Another effect of the dust decoupling is that a significant amount of this larger dust becomes trapped at the cavity edge, instead of following the gas into the cavity and along the streamers to be accreted by the central binary stars.
This arises from the positive pressure gradient located at the cavity, halting the inward drift of dust, allowing it to accumulate.

For Stokes number of 0.1, the faster inwards drift and pile-up of dust around the cavity is even clearer, with the dust now becoming concentrated in an eccentric ring outside the cavity.
Note that like the gas, due to the eccentricity of the ring, the dust surface density is largest at the apocentre, since the orbital velocities are slower there. The dust surface density at the cavity edge has increased by a factor of $\sim 20$ in just 30,000 binary orbits (corresponding to $\sim 3300$~yr), indicating that substantially larger concentrations of dust would be expected over longer run times provided the disc is large enough to provide a continuous source of drifting dust. 

\begin{figure}
    \centering
    \includegraphics[scale=0.6]{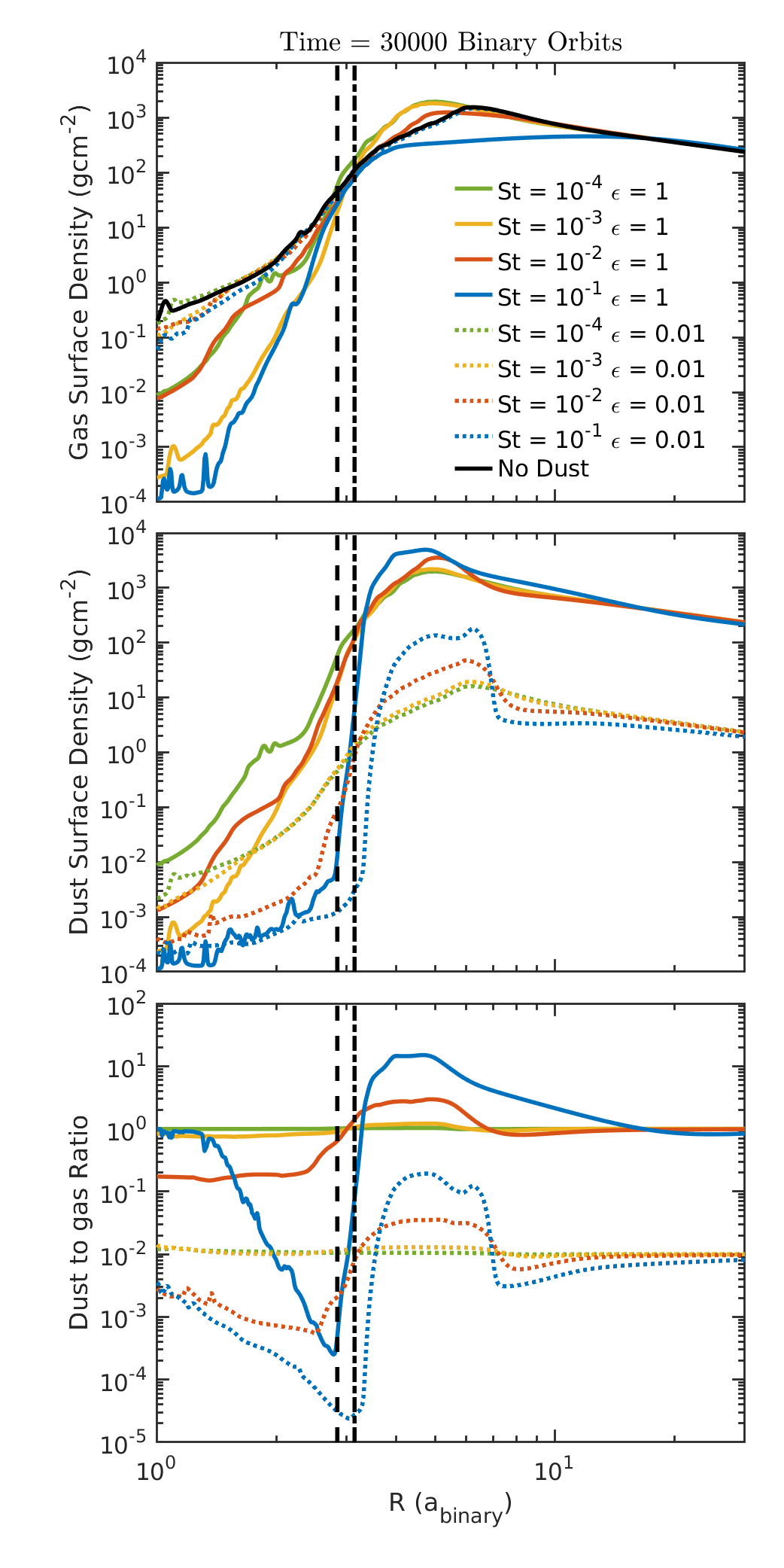}
    \caption{Azimuthally averaged surface densities for gas (top panel) and dust (middle panel) as well as the azimuthally averaged \dtg ratio (bottom panel) for discs around Kepler-16. Coloured lines show represent different Stokes numbers for the dust: blue being Stokes number of 0.1, red being $10^{-2}$, yellow being $10^{-3}$ and green being $10^{-4}$. The Solid black shows the disc where dust was not included. We differentiate between the two \dtg ratios by using solid coloured lines for \dtg ratios of 1, and dotted lines for \dtg ratios of 0.01. The vertical dashed and dot-dashed line represent the stability limit \citep{Holman99} and the observed location of Kepler-16b \citep{Welsh12} respectively.}
    \label{fig:kep-16_sigmas_multi}
\end{figure}

Fig. \ref{fig:kep-16_2d_1} shows the surface densities for discs with \dtg ratios equal to unity.
As expected from fig. \ref{fig:kep-16_cav_sma_ecc} the cavities in all of the discs in fig. \ref{fig:kep-16_2d_1} are smaller than their counterparts in fig. \ref{fig:kep-16_2d_0.01}.
They are also more circular in general as indicated by fig. \ref{fig:kep-16_cav_sma_ecc}.
Whilst there was little change in the gas surface density distributions as a function of Stokes number for the discs with \dtg ratios of 0.01, increasing the Stokes number here leads to significant changes.
For $\Stokes=10^{-4}$ and $10^{-3}$, the gas surface densities are very similar, which is unsurprising as the dust is still well coupled to the gas, even though its abundance relative to the gas is now unity.
This also agrees with what is seen in fig. \ref{fig:kep-16_cav_sma_ecc} where the solid green and yellow lines are very similar, indicating similar cavity structures.
Looking at the 2D dust surface density distributions for these discs, the surface densities are seen to be similar to the gas surface densities (note the colour scales differ between the gas and dust contours).
With the dust being well coupled to the gas, there are few differences between the gas and dust surface densities.

Increasing the Stokes number to $10^{-2}$, leads to the gas having lower density near the cavity. Conservation of angular momentum means that as the dust drifts inwards the gas pressure bumps are smoothed out somewhat.
There is now a clear difference between the gas and dust surface densities.
Even though the \dtg ratio was initially unity everywhere in the disc, the drift of the dust concentrates it around the cavity while smoothing the pressure bump in the gas.

Increasing the Stokes number to 0.1 leads to the greatest effect of the dust on the gas.
The cavity size is smaller and it is more circular than in the other runs considered.
The lack of an over-density around the cavity is also apparent with the gas surface density profiles exterior to the cavity now being flat and smaller compared to the gas surface densities from the other runs.
The dust surface density contours show that a dense ring of dust orbits around the cavity due to the rapid radial drift of this larger dust, which increases the local \dtg ratio to be $\sim 20$.

Whilst figs \ref{fig:kep-16_2d_0.01} and \ref{fig:kep-16_2d_1} show the gas and dust surface densities in 2D, it is also useful to look at the azimuthally averaged values, which can better show features in the disc as a function of orbital distance.
Figure \ref{fig:kep-16_sigmas_multi} shows the azimuthally averaged surface density for the gas (top panel), dust (middle panel) as well as the \dtg ratio (bottom panel) at a time of 30,000 binary orbits.
The colour coding and line formats are the same as those used previously.
Looking at the gas surface densities, the similarities between the disc with no dust (black line) and those with low \dtg ratios (dotted lines) is clear with all the dotted lines sitting beneath the black line.
For the dust surface densities, there are however some differences.
Whilst the green and yellow dotted lines, corresponding to discs with well coupled dust, are roughly two orders of magnitude lower than their gas surface density counterparts, the red and blue lines with less coupled dust show pile-ups around the cavity region.
These build-ups are also seen in the \dtg ratios for the dotted lines with the green and yellow lines maintaining \dtg ratios around 0.01, whilst the red line managed to increase its \dtg ratio to 0.04 around the cavity.
The disc with $\Stokes=0.1$ (blue dotted line) increased the \dtg ratio around the cavity to $\sim0.2$, 20 times larger than the initial value of 0.01.

For the discs with initial \dtg ratios of unity, the solid lines in fig. \ref{fig:kep-16_sigmas_multi}, the changes in gas surface density from the disc with no dust (black line) is clear.
The discs with tightly coupled dust (yellow and green lines) have a similar profile to the disc with no dust, but the cavity is noticeably smaller, as has been discussed previously. This may be due to the fact that a gas-dust mixture acts like a colder gas \citep{LinYoudin2017}, and hence the ability of binary-induced spiral waves to travel large distances in the disc before dissipating due to nonlinear effects may be reduced in these cases.
The dust surface densities for those discs are also similar to the gas surface densities, and this is shown in the bottom panel of fig. \ref{fig:kep-16_sigmas_multi} where the \dtg ratios for the green and yellow solid lines remain roughly unity.
\begin{figure}
    \centering
    \includegraphics[scale=0.4]{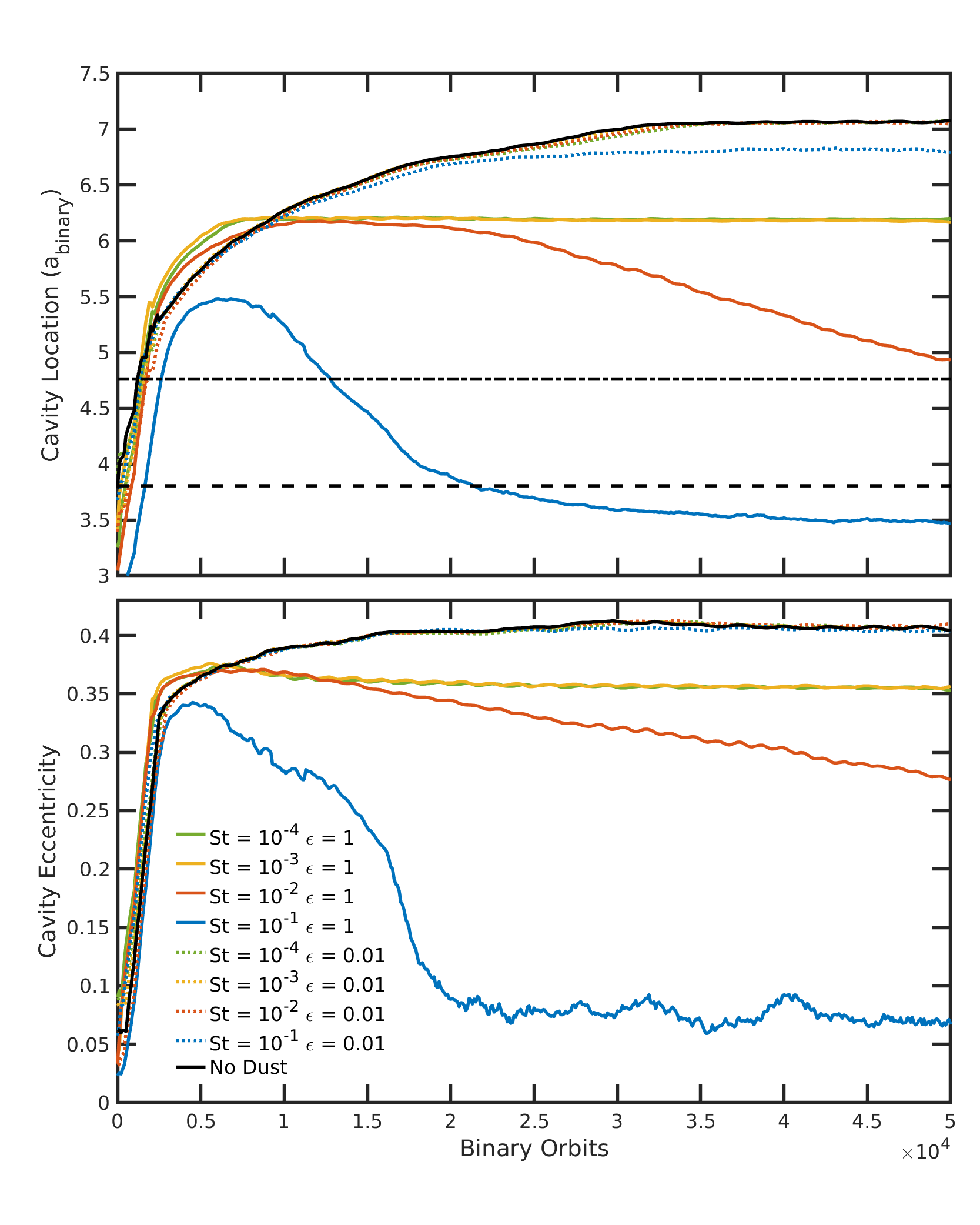}
    \caption{Time evolution of the cavity location (top panel) and eccentricity (bottom panel) for the discs in the Kepler-34 system. Coloured lines represent different Stokes numbers for the dust: blue being Stokes number of 0.1, red being $10^{-2}$, yellow being $10^{-3}$ and green being $10^{-4}$. The Solid black shows the disc where dust was not included. We differentiate between the two \dtg ratios by using solid coloured lines for \dtg ratios of 1, and dotted lines for \dtg ratios of 0.01. The horizontal dashed and dot-dashed line represent the stability limit \citep{Holman99} and the observed location of Kepler-34b \citep{Doyle11} respectively.}
    \label{fig:kep-34_cav_sma_ecc}
\end{figure}
Now looking at the red line for the run with $\Stokes=10^{-2}$, the shape is similar to the black, green and yellow lines, but the peak in the gas surface density is now roughly half the value, whilst the peak in the dust surface density is roughly twice that of the green and yellow lines, highlighting the pile-up of dust around the cavity edge and its subsequent effects on the surrounding gas.
This is also seen in the bottom panel where the \dtg ratio for the red line reaches a peak of $\sim 4$, larger than its initial value of unity.
When increasing the Stokes number to 0.1 (blue solid line), the peak in the dust surface density increases even further, whilst the gas surface density outside the cavity is reduced with the profile now becoming flat.
At the dust peak around the cavity edge, the \dtg ratio reaches $\sim 20$, significantly altering the structure of the cavity and the surrounding regions as shown in figs. \ref{fig:kep-16_cav_sma_ecc}--\ref{fig:kep-16_disc_ecc} as well as the right-hand-most panels of fig. \ref{fig:kep-16_2d_1}.

\subsection{Kepler-34}
\label{sec:kep-34_disc}
We now turn our attention to the discs around the Kepler-34 analogues.
We again examine dust with Stokes numbers between $10^{-4}$ and $10^{-1}$, and \dtg ratios of 0.01 and 1.
The colour coding and line formats in the following figures are the same as their Kepler-16 counterparts in sect.~\ref{sec:kep-16_disc}.

\begin{figure}
    \centering
    \includegraphics[scale=0.6]{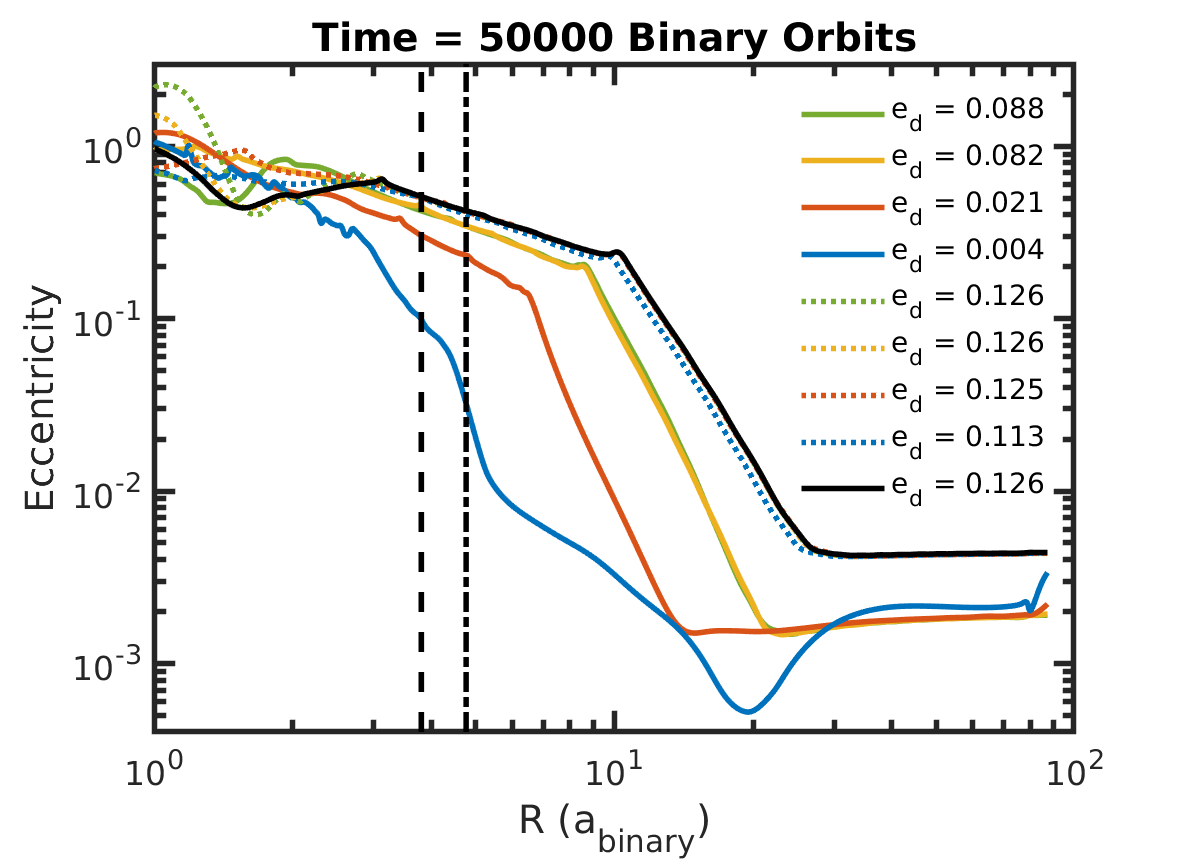}
    \caption{Eccentricity profiles as a function of orbital radius for the discs in the Kepler-34 system. Colours and line formats are the same as in fig. \ref{fig:kep-34_cav_sma_ecc}. The values in the legend denote the time-averaged disc eccentricity up to an orbital distance of 30$\ab$.}
    \label{fig:kep-34_disc_ecc}
\end{figure}

\begin{figure*}
    \centering
    \includegraphics[scale=0.45]{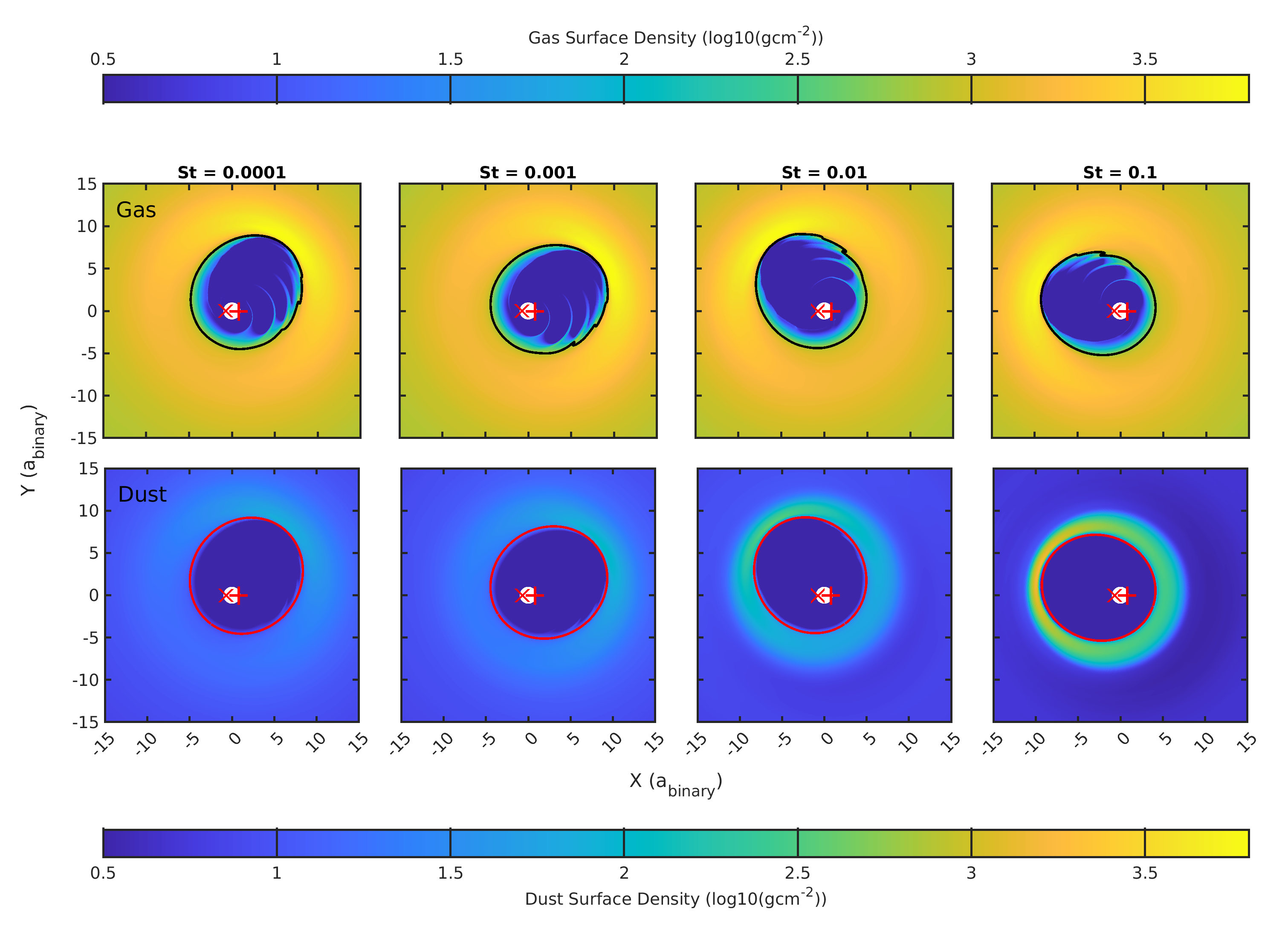}
    \caption{Two-dimensional surface density plots of the gas (top panels) and dust (bottom panels) for the discs in the Kepler-34 system.
    The \dtg ratio for the discs was equal to 0.01. Going from left-to-right we increase the Stokes number of the dust from $10^{-4}$ to $10^{-1}$, shown at the top of each column. The red `x' and `+' denote the locations of the binary stars, with the black line in the top panels showing the rough location of the cavity with a contour of the surface density at a value equal to $10\%$ of the maximum surface density. The red ellipse in the bottom panels shows the gas cavities using eq. \ref{eq:cav_size}.}
    \label{fig:kep-34_2d_0.01}
\end{figure*}

\subsubsection{Cavity properties}
Following eqs. \ref{eq:cav_size} and \ref{eq:cav_ecc}, fig. \ref{fig:kep-34_cav_sma_ecc} shows the cavity semi-major axes (top panel) and eccentricities (bottom panel) for the discs around Kepler-34.
Similar to Kepler-16, the cavity sizes for \dtg ratios of 0.01 follow the black line corresponding to the dust-free disc.
It takes a considerably larger number of binary orbits for the cavity sizes to reach their equilibrium values for the discs around Kepler-34 analogues, roughly 30,000 binary orbits.
Whilst the runs with $\Stokes\le 10^{-2}$ closely follow the dust-free disc with cavity sizes reaching an equilibrium value of $\sim7\ab$, the disc with the most decoupled dust ($\Stokes=0.01$) achieves a slightly smaller cavity of size $\sim 6.8\ab$.
Like the discs in Kepler-16, the cavities around Kepler-34 are eccentric, but even more so than Kepler-16 due to the more eccentric central binary in the Kepler-34 system.
For the low \dtg ratio discs, their cavities achieve eccentricities of just over 0.4, nearly $50\%$ larger than that found for the discs around Kepler-16.

Looking at the discs with \dtg ratios of unity in fig. \ref{fig:kep-34_cav_sma_ecc}, the larger dust abundance leads to reduced cavity sizes and eccentricities, and equilibrium values are achieved after $\sim$ 10,000 binary orbits.
Similar to their counterparts around Kepler-16, the discs with well coupled dust (green and yellow lines) evolve similarly and maintain cavity sizes of $\sim6.2\ab$, and cavity eccentricities of $0.35$.
The red line showing the run with $\Stokes=10^{-2}$ also initially settles with a cavity size of $\sim6.2\ab$, but on longer time scales the interactions between the dust and the gas near the cavity act to circularise the gas whilst maintaining the location of the pericentre close to the stability limit.
This circularisation can also be seen in the red solid line in the bottom panel of fig. \ref{fig:kep-34_cav_sma_ecc} where it consistently falls over time to reach a value of 0.28 after 50,000 binary orbits.
Increasing the Stokes number to 0.1 yields the greatest effect of the dust on the cavity size and eccentricity.
After initially raising the cavity location to $5.5\ab$ and its eccentricity to 0.33, the build-up of dust around the cavity acts to circularise and reduce the cavity size.
After 50,000 binary orbits, the cavity size has decreased and settled at a location of $\sim3.5\ab$, just interior to the stability limit, with an eccentricity of 0.07.

Using eq. \ref{eq:disc_ecc_r} we calculate the azimuthally averaged eccentricities for the discs around Kepler-34.
Figure \ref{fig:kep-34_disc_ecc} shows the eccentricities as a function of orbital radius for the discs with \dtg ratios of 0.01 (dotted lines) and of unity (solid lines), with the black line being for the dust-free disc.
The different colours represent dust with different Stokes numbers, whilst the legend shows the time-averaged disc eccentricity over 5,000 binary orbits out to an orbital radius of $30\ab$.
Similar to the discs around Kepler-16, the eccentricity profiles for the discs around Kepler-34 can be described with three regions: Interior to the cavity edge; Exterior to the cavity; and far away from the binary.
Again, inside the cavity the eccentricities approach 1, reducing to $\sim0.3$ around the cavity apocentre, before falling more steeply outside the cavity region down to where the binary perturbations become negligible at $\sim30\ab$.

Like the discs with low \dtg ratios in Kepler-16, those discs around Kepler-34 also have eccentricity profiles similar to the disc with no dust, with the dotted lines in fig. \ref{fig:kep-34_disc_ecc} sitting beneath the black solid line.
The time-averaged disc eccentricities in the legend for these discs are generally larger than those found in the discs around Kepler-16.
For the disc with no dust, the disc eccentricity is around 0.126, with the majority of the dotted lines being within $1\%$ of this value.
However, for the disc with the largest dust (blue dotted line), the eccentricity drops slightly to 0.113.

\begin{figure*}
    \centering
    \includegraphics[scale=0.45]{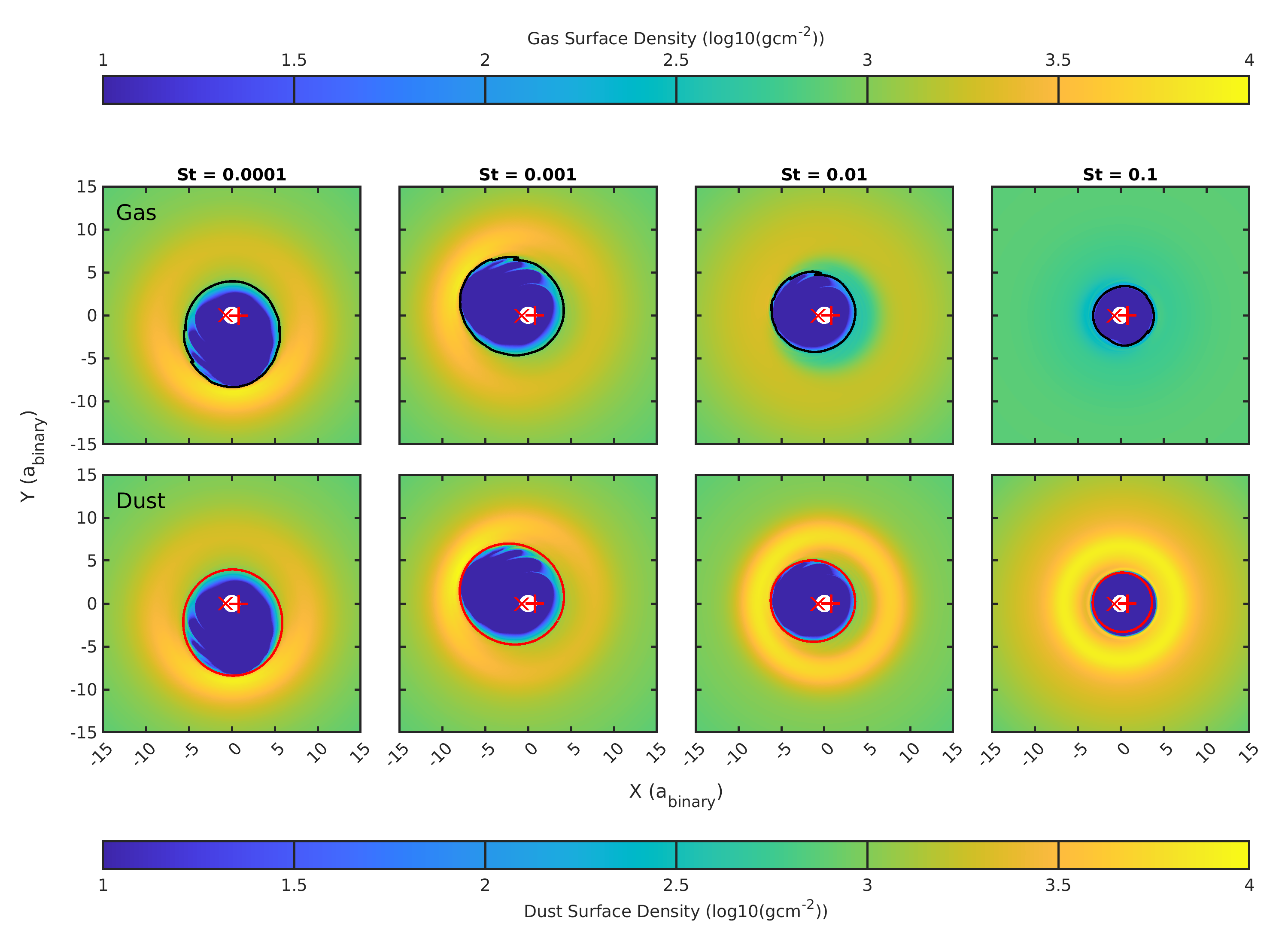}
    \caption{Same as fig. \ref{fig:kep-34_2d_0.01} but for discs with \dtg ratios equal to 1.}
    \label{fig:kep-34_2d_1}
\end{figure*}

When increasing the \dtg ratio to 1, the solid coloured lines in fig. \ref{fig:kep-34_disc_ecc} show that the gas eccentricities as a function of orbital radius have decreased due to the greater effectiveness of the dust-gas interactions.
For the discs with well coupled dust (green and yellow lines), the eccentricities are only mildly reduced compared to the discs with lower dust abundances, with time-averaged disc eccentricities of 0.088 and 0.082 respectively.
The effect of increasing the dust sizes is again apparent here, with the larger Stokes numbers significantly circularising the gas in the vicinity of the cavity.
The runs with $\Stokes=10^{-2}$ (red line) and $10^{-1}$ (blue line) are less eccentric at all radii, with the $\Stokes=10^{-1}$ disc in particular becoming almost circular in places.
This is confirmed by calculating the time-averaged disc eccentricity out to $30\ab$, giving values of 0.021 and 0.004 for the discs with the largest Stokes numbers.
The value of 0.004 for the disc with the largest dust grains ($\Stokes=10^{-1}$) highlights the important role that dust can have in modifying the dynamics of the disc and the structure of the tidally-truncated cavity.

In summary, like for Kepler-16, figs. \ref{fig:kep-34_cav_sma_ecc}--\ref{fig:kep-34_disc_ecc} show that as the Stokes numbers or \dtg ratios increase, the gas cavity locations, their eccentricities, and the inner disc eccentricities all decrease as the dust removes energy and angular momentum from the gas.
This effect, probably arising as a dust-free gas disc is able to establish a well-defined precessing, eccentric mode in proximity of the cavity region that is forced by the binary. Pressure-less dust particles, however, cannot form such a mode and instead achieve moderate eccentricities due to secular forcing by the binary.
The back reaction from the dust on to the gas can then force the gas to become increasingly circularised as the forces increase, i.e. through larger dust abundances or Stokes numbers.

\subsubsection{Surface densities}
Figure \ref{fig:kep-34_2d_0.01} shows 2D surface density plots for the gas (top panels), and dust (bottom panels) for the discs with \dtg ratios of 0.01 around Kepler-34 after 50,000 binary orbits.
The Stokes number of the dust, shown above each of the topmost panels, increases from left-to-right, with the colour schemes showing the logarithm of the gas (top bar) and dust (bottom bar) surface densities.
We denote the locations of the binary stars with the red ‘+’ and ‘x’ in the centre of the panels.
The black and red lines in the top and bottom panels show the location of the cavity using a $10\%$ contour and eq. \ref{eq:cav_ecc} respectively, again showing the quality of the fits to the cavity orbital elements.

The increase in cavity eccentricity for the discs around Kepler-34 compared to Kepler-16 is clear to see.
Like the discs around Kepler-16, the 2D gas surface density profiles for the discs with low \dtg ratios are extremely similar irrespective of the Stokes number.
This agrees with the results shown by the dotted lines figs. \ref{fig:kep-34_cav_sma_ecc}--\ref{fig:kep-34_disc_ecc} where the cavity sizes, eccentricities and radial eccentricity profiles are similar for all those discs.
Only the disc with large dust, $\Stokes=0.1$, has a slightly smaller cavity size.

Looking at the dust surface densities in the bottom panels of fig. \ref{fig:kep-34_2d_0.01}, there are considerable differences as the Stokes number increases.
For $\Stokes \le10^{-3}$, the dust morphologies are similar to their gas counterparts, albeit with surface densities two orders of magnitude lower. 
Here, the dust is well coupled to the gas, and so the similarity in disc structures is unsurprising.
Looking at the bottom panel with $\Stokes=10^{-2}$, the dust is concentrating around the cavity edge, where there is a strong pressure bump, allowing these larger dust grains to become trapped as they drift inwards.
Increasing the Stokes number to 0.1, the bottom right panel of fig. \ref{fig:kep-34_2d_0.01} shows a significant concentration of dust around the cavity region.
The concentration here reaches a maximum \dtg ratio of $\sim0.2$ even though the initial \dtg ratio was equal to 0.01.
The accumulation of dust at the apocentre is much more visible here with an order of magnitude difference between the surface densities at apocentre and pericentre.
This is due to the increased eccentricity induced by the binary compared to that seen in the discs around the Kepler-16 analogues.

Increasing the initial \dtg ratio to 1, fig. \ref{fig:kep-34_2d_1} shows the 2D surface density plots for the gas (top panel) and dust (bottom panel) for the circumbinary discs after 50,000 binary orbits.
The effect on the cavity sizes and eccentricities is apparent in all of the panels, highlighting the impact the dust has on the cavity structures as shown by the solid lines in fig. \ref{fig:kep-34_cav_sma_ecc}.
Looking at the runs with $\Stokes\le10^{-3}$, the cavity regions are noticeably smaller than their counterparts in fig. \ref{fig:kep-34_2d_0.01}, but again the dust and gas surface densities are similar.
Increasing the Stokes number allows the dust to significantly modify the cavity structure.
For $\Stokes=10^{-2}$, a large dust ring forms outside the cavity, acting to flatten the gas profile.
There is a significant decrease in the maximum surface density around the cavity region.
The cavity is also much smaller and circular as seen in fig. \ref{fig:kep-34_cav_sma_ecc}.
For $\Stokes = 10^{-1}$ the gas surface density becomes very flat, with no observable maximum at the cavity edge.
The cavity's small size and lack of eccentricity is also apparent, with it being almost half the size of the cavity shown in the leftmost panel of fig. \ref{fig:kep-34_2d_1}.
Looking at the dust surface density plot, the ring of dust outside the cavity is very pronounced and circular.
As seen for the corresponding disc around Kepler-16, the \dtg ratio at the peak of the dust ring reaches a value of $\sim20$.

\begin{figure}
    \centering
    \includegraphics[scale=0.6]{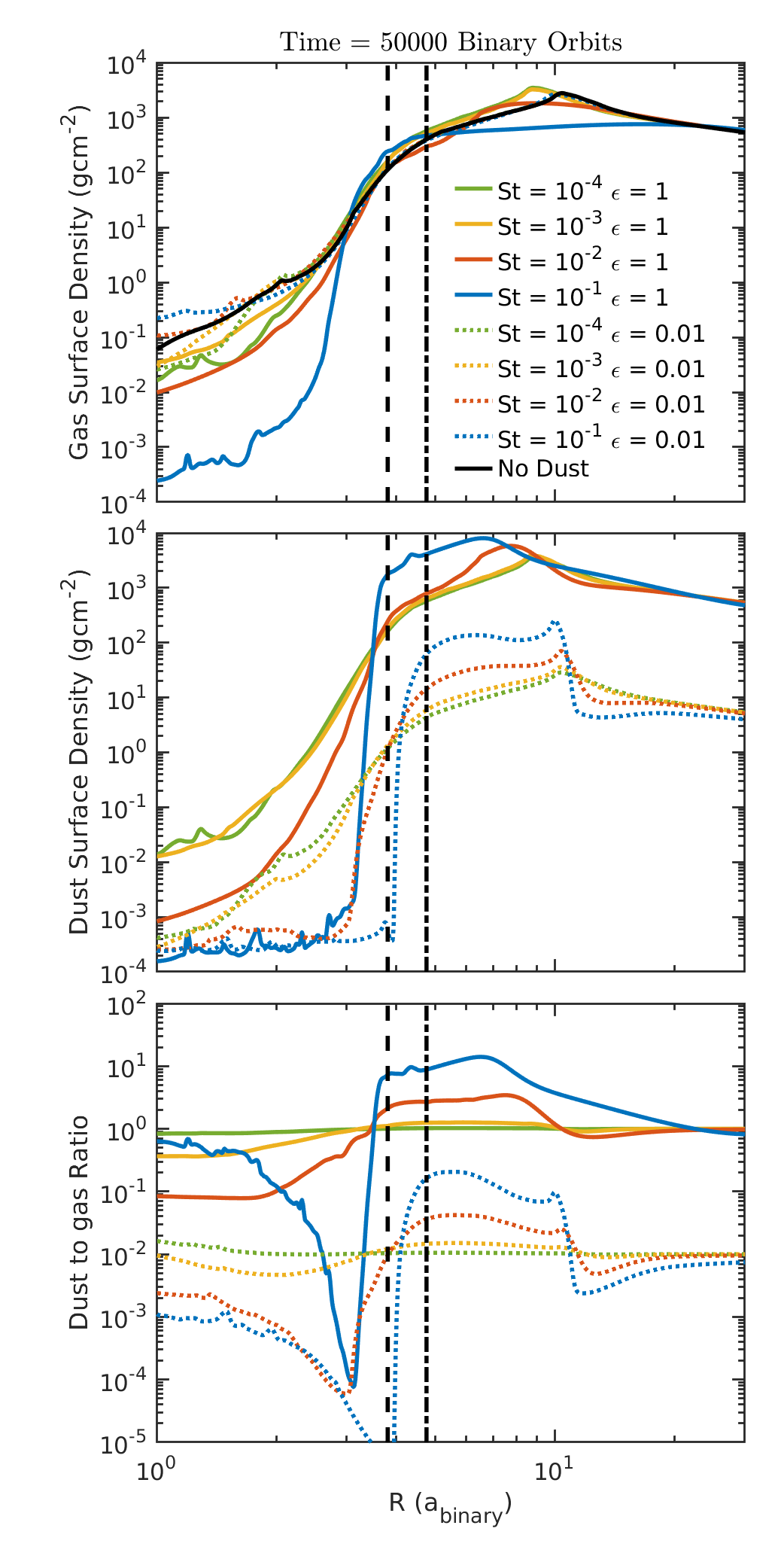}
    \caption{Azimuthally averaged surface densities for gas (top panel) and dust (middle panel) as well as the azimuthally averaged \dtg ratio (bottom panel) for discs around Kepler-34. Coloured lines show represent different Stokes numbers for the dust: blue being Stokes number of 0.1, red being $10^{-2}$, yellow being $10^{-3}$ and green being $10^{-4}$. The Solid black shows the disc where dust was not included. We differentiate between the two \dtg ratios by using solid coloured lines for \dtg ratios of 1, and dotted lines for \dtg ratios of 0.01. The horizontal dashed and dot-dashed line represent the stability limit \citep{Holman99} and the observed location of Kepler-34b \citep{Welsh12} respectively.}
    \label{fig:kep-34_sigmas_multi}
\end{figure}

Fig. \ref{fig:kep-34_sigmas_multi} shows the azimuthally averaged gas (top panel) and dust (middle panel) surface densities as well as the \dtg ratios (bottom panel).
As was seen in fig. \ref{fig:kep-16_sigmas_multi} for Kepler-16, the gas surface densities for the discs with low \dtg ratios (dotted lines) are similar to the black line showing the dust-free disc.
The dotted lines in the middle panel again show the dust concentrating in and around the cavity for the larger values of $\Stokes$.
This is reflected in the \dtg ratios, where only the discs with the larger dust particles show significant deviations from the initial value of 0.01, reaching a maximum of 0.05 and 0.2 for  $\Stokes=10^{-2}$ and $10^{-1}$, respectively.

The solid coloured lines in fig. \ref{fig:kep-34_sigmas_multi} show the 1D profiles for discs with initial \dtg ratios of unity.
The peak in the gas density has moved closer to the star as the cavity sizes have decreased, whilst the peaks in the dust surface densities have also moved inward.
The flattening of the discs with large dust grains is visible for the runs with larger values of $\Stokes$ (red and blue lines).
This is especially the case for the blue line showing the disc with $\Stokes=10^{-1}$, where the lack of a perceivable bump at the cavity apocentre is apparent.
The increase in the \dtg ratios for these discs is also apparent in the bottom panel where the peak in the blue line reaches a \dtg ratio of $\sim20$, similar to what was seen in the equivalent run around Kepler-16.
It is also interesting to note that as seen in fig. \ref{fig:kep-34_cav_sma_ecc} and the solid blue lines in fig. \ref{fig:kep-34_sigmas_multi}, that the cavity location is positioned interior to the stability limit.
Should planets migrate and stop at the cavity as predicted in previous works \citep{Pierens07,Pierens08b,Thun18}, then they would be located around the stability limit, possibly on unstable orbits over long time-scales.

\section{Effects on planet migration}
\label{sec:planet_evol}
We now add a planet into each simulation to explore how the inclusion of dust, and its effect on disc structure, affects the migration of planets within the circumbinary discs.
The final orbital properties of the planets can then be compared with the observed values for Kepler-16b and -34b, as well as with the results from previous works investigating these systems.
For both systems, we place a $20\me$ planet with zero eccentricity into the settled discs at large distance from the cavities.

\begin{figure}
    \centering
    \includegraphics[scale=0.4]{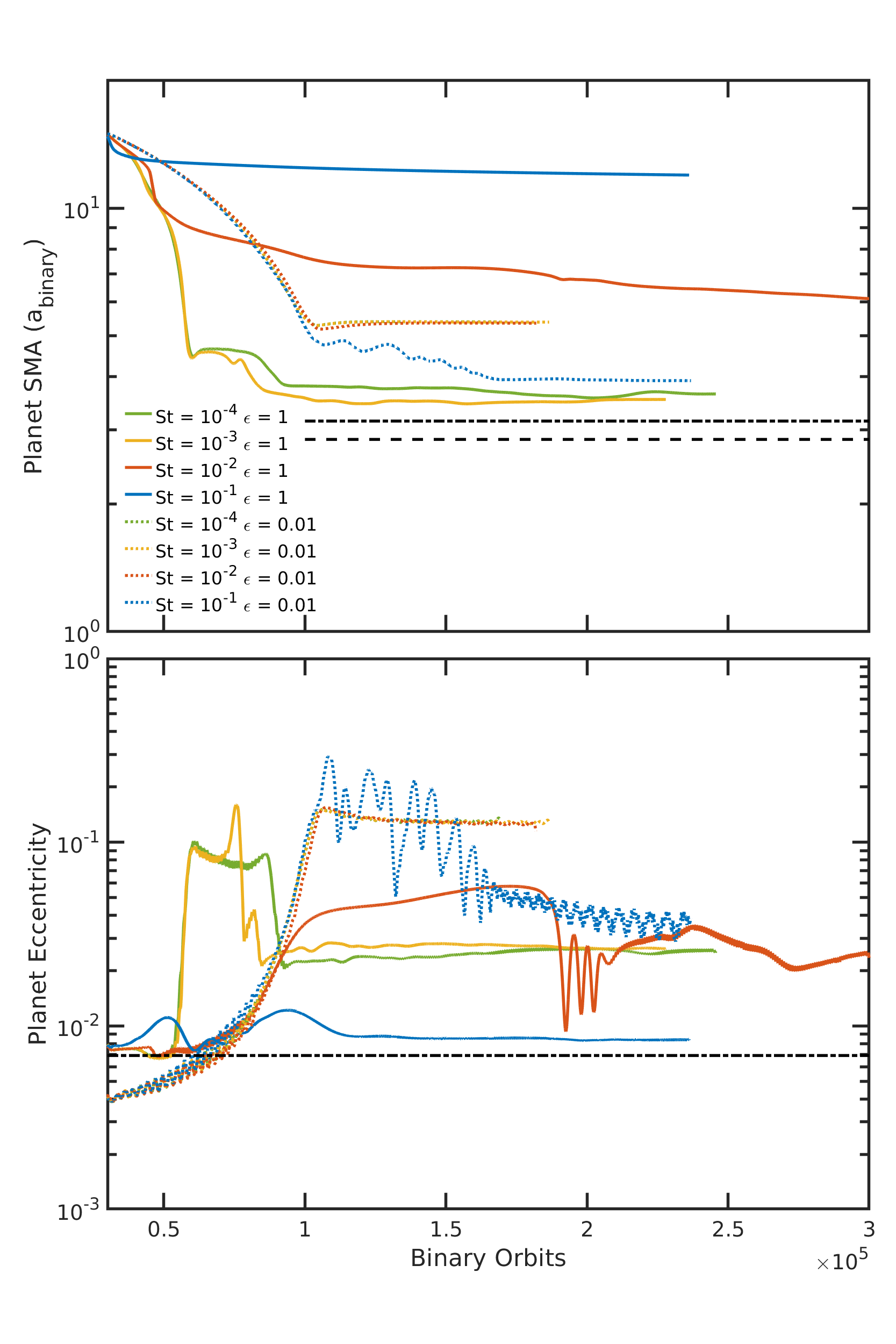}
    \caption{Time evolution for planet semi-major axes (top panel) and eccentricities (bottom panel) for planets around Kepler-16. Coloured lines represent different Stokes numbers for the dust: blue being Stokes number of 0.1, red being $10^{-2}$, yellow being $10^{-3}$ and green being $10^{-4}$. We differentiate between the two \dtg ratios by using solid coloured lines for \dtg ratios of 1, and dotted lines for \dtg ratios of 0.01. The horizontal dashed and dot-dashed line represent the stability limit \citep{Holman99} and the observed location of Kepler-16b \citep{Doyle11} respectively.}
    \label{fig:kep-16_planet}
\end{figure}

\subsection{Kepler-16}
\label{sec:planets_kep16}
First, we consider the Kepler-16 analogues where we placed the $20\me$ planet into the disc after 30,000 binary orbits, at an orbital distance of $15\ab$. Note that this is not the observed mass of Kepler-16b \citep[see table~\ref{tab:systems} and][]{Doyle11, Triaud21}, but instead is supposed to represent the mass of the planet while it was still forming and is below the mass at which runaway gas accretion would have occurred). 
In fig. \ref{fig:kep-16_planet} we show the evolution of the planet semi-major axes and eccentricities.
The colour coding and line formats are the same as those used in previous plots in regards to the Stokes number and \dtg ratio for each specific simulation.

The planets in the discs with low \dtg ratios slowly migrate inwards, and after $\sim100,000$ binary orbits they reach the outer periphery of the cavity region.
Interactions between the planets and the gas and dust slows the migration.
For quasi-circular orbits, strong positive corotation torques counter the Lindblad torques acting on the planets, due to the positive surface density profile in the cavity.
For more eccentric planets, an alternative torque cancellation mechanism can occur when a planet orbits at the edge of a cavity \citep[see the discussion in][]{Pierens13}.
We see that in general the eccentricities remain small and so the former effect is dominant.
Corotation torques are enhanced by the eccentric orbits of the planets aligning with that of the cavity.
Both planets and the discs attain very similar longitudes of pericentres and precess together.
As the planets have migrated, they have also crossed numerous mean-motion resonances (MMRs).
These $n$:1 MMR locations have been shown to be unstable to planetary orbits as they excite eccentricities, and can eventually scatter planets out of the disc \citep{Nelson2000}.
Since the interactions with the disc cause the planets to migrate quickly past the resonances, they are able to maintain stable orbits similar to that seen in previous studies \citep{Kley14,Mutter17P}.

\begin{figure}
    \centering
    \includegraphics[scale=0.4]{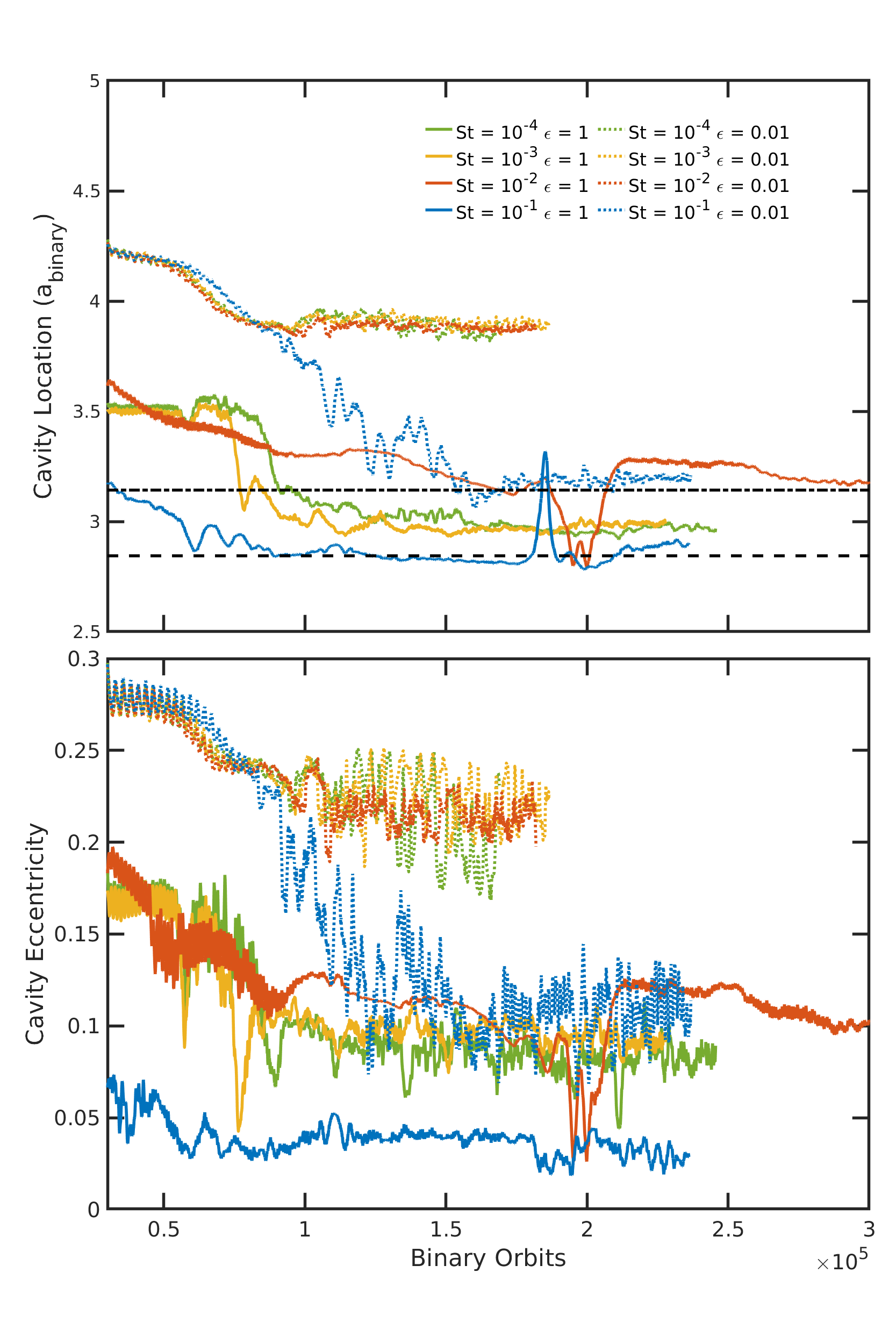}
    \caption{Time evolution of the cavity location (top panel) and eccentricity (bottom panel) for discs in the Kepler-16 system including migrating planets. Colours and line formats are the same as in fig. \ref{fig:kep-16_cav_sma_ecc}.}
    \label{fig:kep-16_planet_disc}
\end{figure}

With most of the dotted lines in fig. \ref{fig:kep-16_planet} evolving in a similar manner, the effects of changing the dust size is moderate, and the final stopping locations of the planets are similar to those obtained by \cite{Pierens13} and \cite{Mutter17P}.
The exception, however, is the disc with the largest dust (blue dotted line), in which the planet has migrated somewhat closer to the central binary. As noted earlier, for the \dtg ratio of 0.01, the presence of dust does not noticeably affect the cavity size compared to the dust-free case when no planet is present, even for $\Stokes = 10^{-1}$, but it can change the nature of the interaction between the disc and planet by effectively making the gas act as if it was a colder fluid. This is what seems to be happening in this case: the build-up of dust at the cavity edge allows the disc's response to the presence of the planet to behave more non-linearly. The planet is then able to modify the disc structure there, effectively forming its own gap, allowing the planet to migrate closer to the binary. A similar phenomenon has been observed in previous simulations where the mass of the planet significantly affects its stopping location because of nonlinear effects \citep{Pierens13,Mutter17P,Penzlin21}, implying that Kepler-16b may have undergone type II migration during the final stages of the disc lifetime.

Figure \ref{fig:kep-16_planet_disc} shows the evolution of the cavity size and eccentricity for these discs when the planets are present. It is clear the planet strongly affects the cavity for the case with \dtg equal to 0.01 and $\Stokes=10^{-1}$ (dotted blue line), with the shrinkage and partial circularisation of the cavity being due to the planet forming its own gap and migrating further inwards before halting.
Whilst all of the cavities shrunk in size, it is the disc with the largest dust that experienced the greatest decrease (down to $3.5\ab$), and this is because of the build-up of dust in this case. In terms of the stopping locations of the planets in these dust-poor discs, they orbit further out at between 3.9--5.5$\ab$, than Kepler-16b \citep[$3.14\ab$,][]{Doyle11}, as well as having larger eccentricities that the osculating eccentricity seen in observations.
Nonetheless, the simulations show that the build up of dust near the cavity edge can significantly modify the parking locations of migrating planets, and it should be remembered that the planet mass used in the simulations is smaller than that of Kepler-16b. Our aim in this work is not to match the orbital parameters of Kepler-16b and Kepler-34b, but instead to examine the influence of the dust on planet parking locations using uniform initial conditions.

Whilst the evolution of the planets in the discs with low \dtg ratios is broadly similar, this is not the case for the planets in discs with \dtg ratios of unity, as shown by the solid lines in fig. \ref{fig:kep-16_planet}.
For these planets, the abundance of dust initially increases the speed of migration as there is more mass for the planets to exchange angular momentum with and again the gas acts as a colder fluid in the presence of a significant dust abundance.
After the initial stage of migration, the migration rates of the planets change depending on the size of the dust.
For the discs with small dust, $\Stokes = 10^{-3}$--$10^{-4}$, these planets continue migrating towards the cavity region, and show evidence of undergoing a period of runaway or Type III migration \citep{Masset-Snellgrove,Masset-Papaloizou}.
Once they reach the cavity region, they align their orbits with the cavity, allowing the torques acting on the planets to cancel out.
When the planets arrive near the cavity they open moderate gaps in the disc and migrate closer to the stars, seen at $\sim80,000$ binary orbits for the green and yellow lines in fig. \ref{fig:kep-16_planet}.
The planets also reduce the size and eccentricity of the cavities as shown by the green and yellow solid lines in fig. \ref{fig:kep-16_planet_disc}, that shows the evolution of the cavity locations and eccentricities over time.
During the initial migration of the planets, their eccentricities rose to $\sim0.1$, before then dropping to $\sim0.02$--$0.03$ once they have migrated in closer to the central stars.
These values for the planet locations as well as their eccentricities are close to the observed value of Kepler-16b \citep{Doyle11}. These simulations illustrate what can happen if there is a build-up of dust grains with $10^{-4} \le \Stokes \le 10^{-3}$ near the cavity of a circumbinary disc.

For the disc with $\Stokes=10^{-2}$, the red line in fig. \ref{fig:kep-16_planet}, the planet begins to migrate in towards the central stars at a similar migration speed to the discs with smaller Stokes numbers.
However, with the dust being less coupled to the gas, the planet is able to form a gap in the dust disc.
This occurs after $\sim45,000$ binary orbits, where the planet enters a brief period of fast migration as it opens the gap in the dust, as well as a shallower gap in the gas.
The planet then begins to migrate at a slower rate as the net orbit averaged torques from the gas and dust components are reduced.
After 300,000 binary orbits, the planet was still migrating in slowly towards the cavity region, whilst interactions with the material trapped around cavity was able to induce variability within the planets migration rate, as well as its eccentricity, as seen by the irregular behaviour in the re line in fig. \ref{fig:kep-16_planet}.
We stopped the evolution of the disc and the planet after 300,000 orbits to avoid unnecessarily long runtimes.

The blue solid line in fig. \ref{fig:kep-16_planet}, corresponding to the simulation with $\Stokes=10^{-1}$, shows the planet migrates quickly at first while it opens a deep gap in the dust, after which it enters a period of slow migration due to the net torques from the gas and dust components being reduced compared to their initial values.
This period of slow type II migration leaves the planet orbiting far from the cavity as we stopped the simulations due to unfeasibly long runtimes to allow the planet to migrate fully to the cavity region.
The orbital location of the planet is therefore not a good match of Kepler-16b, however, with the planet opening a gap in the disc it was able to attain extremely small eccentricities, compatible with that observed for Kepler-16b.

\begin{figure}
    \centering
    \includegraphics[scale=0.4]{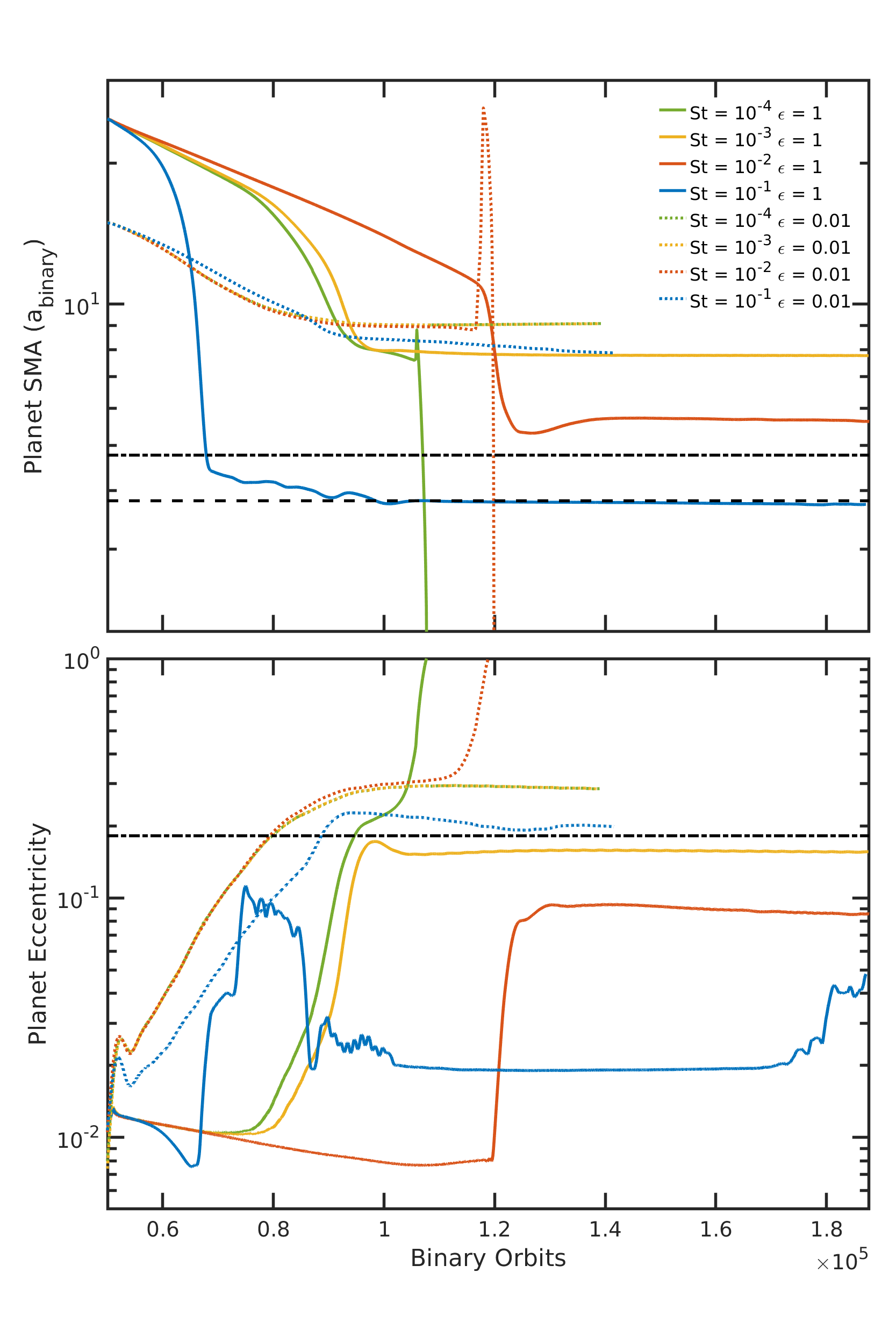}
    \caption{Time evolution for planet semi-major axes (top panel) and eccentricities (bottom panel) for planets around Kepler-34. Coloured lines represent different Stokes numbers for the dust: blue being Stokes number of 0.1, red being $10^{-2}$, yellow being $10^{-3}$ and green being $10^{-4}$. We differentiate between the two \dtg ratios by using solid coloured lines for \dtg ratios of 1, and dotted lines for \dtg ratios of 0.01. The horizontal dashed and dot-dashed line represent the stability limit \citep{Holman99} and the observed location of Kepler-34b \citep{Welsh12} respectively.}
    \label{fig:kep-34_planet}
\end{figure}

\subsection{Kepler-34}
\label{sec:planets_kep34}
We now look at the Kepler-34 analogues, where we insert a $20\me$ planet into the discs after 50,000 binary orbits, at an orbital distance of $15\ab$ for discs with low \dtg ratios and $25\ab$ for discs with \dtg ratios of unity.
The choice of injection location was to make simulation times feasible, whilst also making sure the planet was quite far outside of the cavity region.

In fig. \ref{fig:kep-34_planet} we show the time evolution of the planet semi-major axes (top panel) and eccentricities (bottom panels).
For the low \dtg ratio runs (dotted lines), the planets migrate in slowly and reach the cavity apocentre after $\sim80,000$ binary orbits.
Once near the cavity apocentre, migration slows and the orbits align with the cavity.
The positive corotation torques nullify the Lindblad torques, and the planets settle into orbits with semi-major axes close to the cavity apocentres at $\sim8$--9$\ab$, and eccentricities around 0.2--0.3, both quantities significantly larger than than the observed values for Kepler-34b (see table~\ref{tab:systems}). 

\begin{figure}
    \centering
    \includegraphics[scale=0.4]{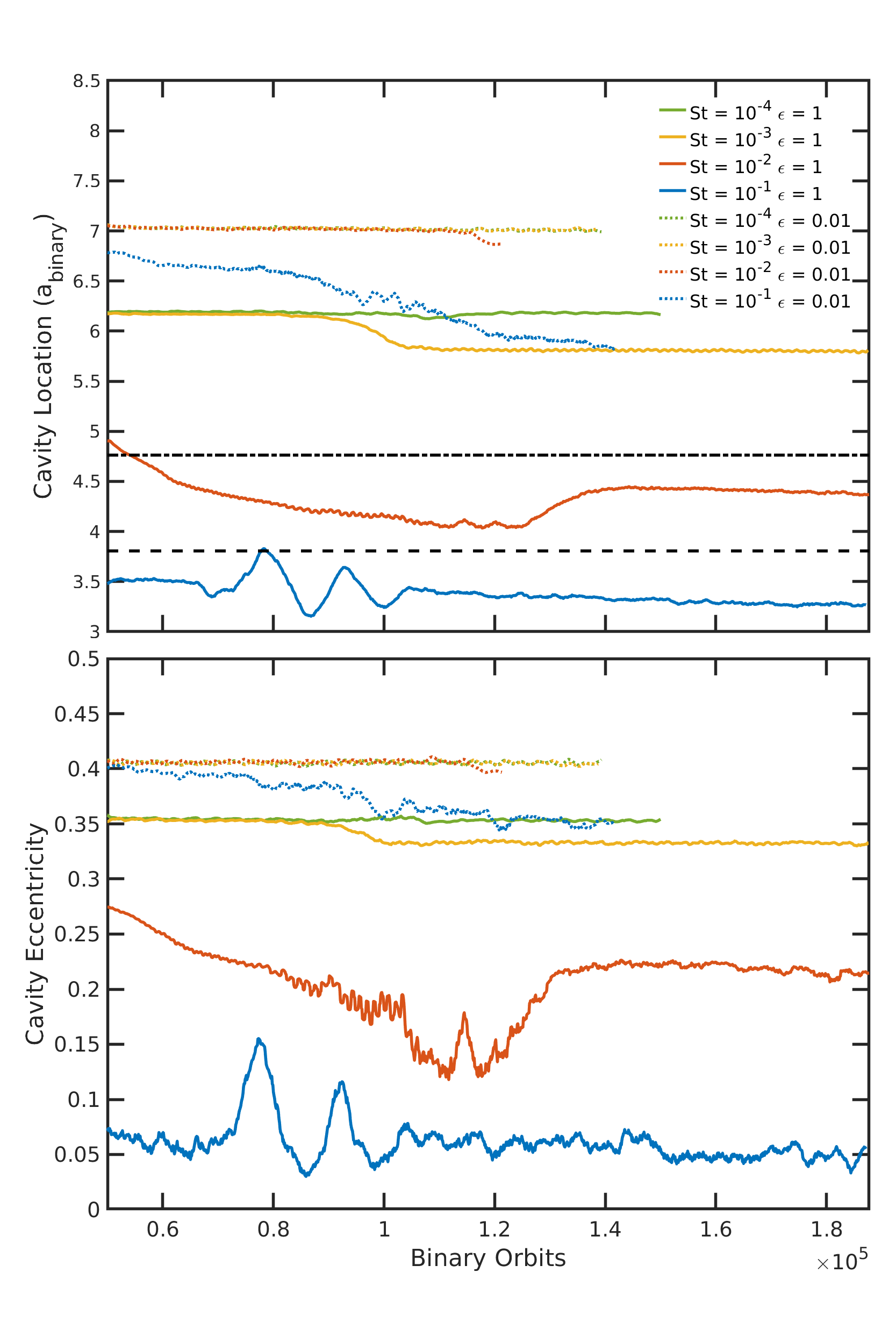}
    \caption{Time evolution of the cavity location (top panel) and eccentricity (bottom panel) for discs in the Kepler-34 system including migrating planets. Colours and line formats are the same as in fig. \ref{fig:kep-16_cav_sma_ecc}.}
    \label{fig:kep-34_planet_disc}
\end{figure}

Apart from the long-term evolution of the run with $\Stokes=10^{-2}$, for which the planet is eventually ejected (we discuss the reasons for this in sect. \ref{sec:ejections}), there appear to be only minor differences in the results when comparing the runs with different dust sizes/Stokes numbers.
The stopping location is always at the outer edge of the cavity. We recall that fig. \ref{fig:kep-34_cav_sma_ecc} shows that the disc with $\Stokes=10^{-1}$ has a slightly smaller cavity size, allowing the planet in that case to migrate slightly closer to the central stars.
Nonlinear interactions between the disc and the planet in this case allow the planet to modify the cavity structure and hence to decrease the cavity size and eccentricity, as shown by the evolution of the dotted blue line in fig. \ref{fig:kep-34_planet_disc}, that shows the evolution of the cavity sizes (top panel) and eccentricities (bottom panel). As with the equivalent run for Kepler-16, this arises because of the build-up of dust at the edge of the cavity.

The runs with \dtg ratios equal to unity are shown by the solid lines in fig. \ref{fig:kep-34_planet}, and we see the planets migrate in towards the cavity at different speeds.
For the disc with $\Stokes=10^{-1}$ (blue line), the planet undergoes runaway migration, and close inspection of the surface density profiles during this phase show the expected morphology of the gas and dust in the corotation region, where the region behind the planet forms a "bubble" of trapped material that exerts a strong negative torque \citep{Peplinski2008}.
The planet halts its migration at the outer edge of the cavity, that as shown in fig. \ref{fig:kep-34_cav_sma_ecc}, is located just interior to the stability limit, just outside the 7:1 MMR.
Similar to planets migrating in Kepler-16, this planet has been able to migrate quickly through the $n$:1 resonances without significantly its eccentricity and undergoing a possible scattering event.
Nonlinear interactions between the planet and disc modify the cavity location and eccentricity as can be seen by the blue solid line in fig. \ref{fig:kep-34_planet_disc}, allowing the final orbit of the planet to be closer to the central binary than is observed for Kepler-34b \citep{Welsh12}.
It is noteworthy that this is a significantly smaller orbit than has been achieved in previous works that ignore the dust component of circumbinary discs \citep[e.g.][]{Pierens13,Thun18,Penzlin21}, for which the parking of Kepler-34b analogues on orbits with semi-major axes much larger than observed for Kepler-34b has always been a problem \footnote{\citet{Mutter17P} also achieved a similarly small stopping location by considering massive self-gravitating disc that resulted in smaller cavity sizes.}. 
Our results indicate that the concentration of dust at the cavity edge can significantly change the cavity structure and the parking location of planets, and may help to explain the final orbits of circumbinary planets such as Kepler-34b.

Looking at the other solid lines in figs. \ref{fig:kep-34_planet} and \ref{fig:kep-34_planet_disc}, showing the discs with smaller Stokes numbers, the planets migrate in towards the cavity at a slower rate.
The planets in the discs with the smallest dust, green and yellow lines, reach the outer cavity edge after $\sim95,000$ binary orbits, where they settle into an orbit aligned with the cavity.
The planet in the disc with the smallest dust is quickly ejected from the system through interactions with the central stars after having its eccentricity excited by the disc.
In sect. \ref{sec:ejections}, we will examine the causes of this ejection and compare its evolution to the planet shown by the solid yellow line, as the disc conditions and evolution, as well as the initial evolution of both of the planets, were similar.

Migrating more slowly than the other planets, before undergoing a phase of runaway migration when it reaches a region of higher dust concentration (see figs. \ref{fig:kep-34_2d_1} and \ref{fig:kep-34_sigmas_multi}), the planet shown by the red solid line in fig. \ref{fig:kep-34_planet} reaches the periphery of the cavity region after $\sim115,000$ binary orbits.
The planet then settles into an orbit at the cavity edge, with an eccentricity of 0.09, a factor of two smaller than Kepler-34b, whilst also orbiting slightly further from the central binary. Nonetheless, this run also illustrates the important role that a concentration of dust at the cavity edge can have on the final parking location of a circumbinary planet.

In all of the cases above, the qualitative behaviour of the planets was the same, where their migration eventually halted at the outer edge of the cavities, with their orbits eventually aligning with the respective cavities.
For planets whose final orbital separations are comparable to that observed for Kepler-34b, the discs were significantly affected by both the abundance and the size of the dust grains.
These interactions caused the gas cavities to be smaller in size and less eccentric, allowing the planets to migrate closer to the central stars, and to the observed location of Kepler-34b.

\subsection{Ejections}
\label{sec:ejections}
In sect. \ref{sec:planets_kep34} we noted that two of the planets were ejected from their systems after interacting with the central stars once their eccentricities became sufficiently large.
The red dotted and the green solid lines in fig. \ref{fig:kep-34_planet} show these planets were able to migrate to the edge of the cavity, similar to the planets that were not ejected.
At the edge of the cavity, all of the planets closely aligned their orbits with the disc, resulting in the apocentres of the planets and the disc being located at very similar azimuthal angles, and with the precession rates of the disc and planets also being very similar. Hence, the conditions are set where a secular resonance can arise between the discs and planets, possibly leading to significant excitation of the eccentricity of the planets, and to their subsequent ejection.
The question arises as to why only some, and not all of the planets got ejected from their respective systems, given the apparent similarity of their evolution as they arrive at the cavity.

To investigate this, we now compare the evolution of the planets in the discs with \dtg ratios of unity, and $\Stokes=10^{-3}$ and $10^{-4}$, the solid yellow and green lines in fig. \ref{fig:kep-34_planet}, which had one planet remain in a stable orbit around the cavity, with the other being ejected.
Note the planet in the disc with \dtg ratio of 0.01 and dust of $\Stokes=10^{-2}$ (red dotted line in fig. \ref{fig:kep-34_planet}) followed a similar evolution to the case we study here.
As was shown in figs. \ref{fig:kep-34_2d_1} and \ref{fig:kep-34_sigmas_multi}, the surface densities of the gas and dust in both discs are similar, both in 2D as well as  the azimuthally averaged profiles.
With similar disc profiles, the forces acting on the planets around their orbits should be similar, with any differences arising from the orbital properties of the planets themselves.
Indeed, it can be seen in fig. \ref{fig:kep-34_planet} that the green solid line migrated inwards slightly faster, and attained a slightly larger eccentricity when it reached the cavity.

In calculating the perturbing forces acting on the planets we follow \citet{MurrayDermott} and define the time derivatives of the semi-major axes and eccentricities resulting from small disturbing forces according to
\begin{equation}
\label{eq:dadt}
    \dfrac{da}{dt} = 2\dfrac{a^{3/2}}{\mu (1-e^2)}[\bar{R}e\sin f +\bar{T}(1+e\cos f)],
\end{equation}
where $\bar{R}$ and $\bar{T}$ are the magnitudes of the radial and tangential components of the perturbing forces, $f$ is the true anomaly of the planet, and $\mu$ is equal to $Gm_{\rm bin}$, and
\begin{equation}
\label{eq:dedt}
    \dfrac{de}{dt} = \sqrt{a\mu^{-1}(1-e^2)}[\bar{R}\sin f + \bar{T}(\cos f + \cos E)]
\end{equation}
where $E$ is the eccentric anomaly.
To calculate the radial and tangential forces from the simulations we use
\begin{equation}
\label{eq:radial}
    \bar{R} = \dfrac{x_{\rm p}a_x + y_{\rm p}a_y}{r_{\rm p}},
\end{equation}
\begin{equation}
\label{eq:tangential}
    \bar{T} = \dfrac{x_{\rm p}a_y - y_{\rm p}a_x}{r_{\rm p}}.
\end{equation}
Then by using eqs. \ref{eq:dadt}--\ref{eq:tangential}, we calculate the evolution of the semi-major axis and eccentricity arising from the disc and from the central stars separately, as well as them acting in concert.
Isolating the forces allows us to determine whether the disc or the central stars are driving up the eccentricities.
To compare with the eccentricity observed in the full simulation we select an interval of time before the planets were ejected, and then evolve those eccentricities by integrating eq.~\ref{eq:dedt} using the disc and N-body forces.

\begin{figure}
    \centering
    \includegraphics[scale=0.6]{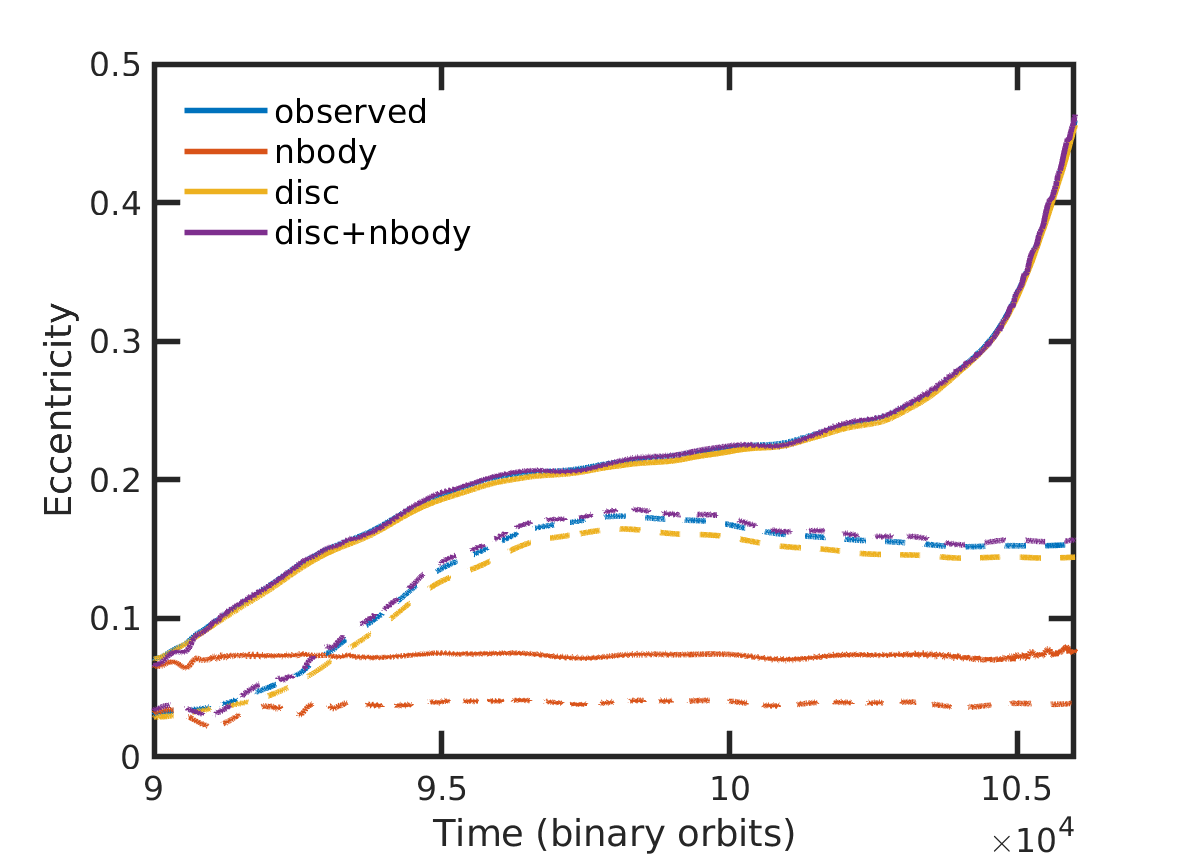}\\
    \caption{Time evolution of planet eccentricities for the ejected and non-ejected planets studied in sect. \ref{sec:ejections}. Solid lines show the ejected planet whilst dashed lines refer to the planet that was not ejected. The colours refer to the observed eccentricity (blue), eccentricities where only N-body (red), disc (yellow), disc and N-body (purple), effects are included.}
    \label{fig:eject_ecc}
\end{figure}

In fig. \ref{fig:eject_ecc} we show the time evolution of the eccentricities observed in the full simulation (blue lines), and the calculated eccentricities including: disc forces (yellow), N-body forces (red) and both disc and N-body forces (purple).
The solid lines show the case where the planet was eventually ejected, and the dashed lines show the case where the planet settled on a stable orbit.
The red lines show the effect of the N-body forces only, and both the ejected and stable cases would have had very little change in their eccentricities if there were only interactions with the central stars.
This is unsurprising given that the planets orbit with small eccentricities far from the stability limit as can be seen by the dashed black line in fig. \ref{fig:kep-34_planet}.

The yellow lines in fig. \ref{fig:eject_ecc} show the influence of the disc forces only, and the similarity between these and the blue and purple lines demonstrates that it is the interaction between the disc and the planet that is primarily responsible for driving the eccentricity evolution.
Ideally, the purple lines would completely overlap the blue lines, corresponding to the calculated eccentricities matching perfectly the observed eccentricities.
However this is not the case here due to the forces from the hydrodynamical simulations only being output every binary orbit, which for these planets gives approximately 20 points per orbit.
In appendix \ref{sec:app_eject}, we reran the simulation for a short period time, outputting planet position and velocities at each timestep.
This allowed for a much more accurate evolution of the eccentricity as can be seen in fig. \ref{fig:eject_appendix} where the residuals are shown in the top panel, corresponding to an error of less than 0.01$\%$.

\begin{figure}
    \centering
    \includegraphics[scale=0.6]{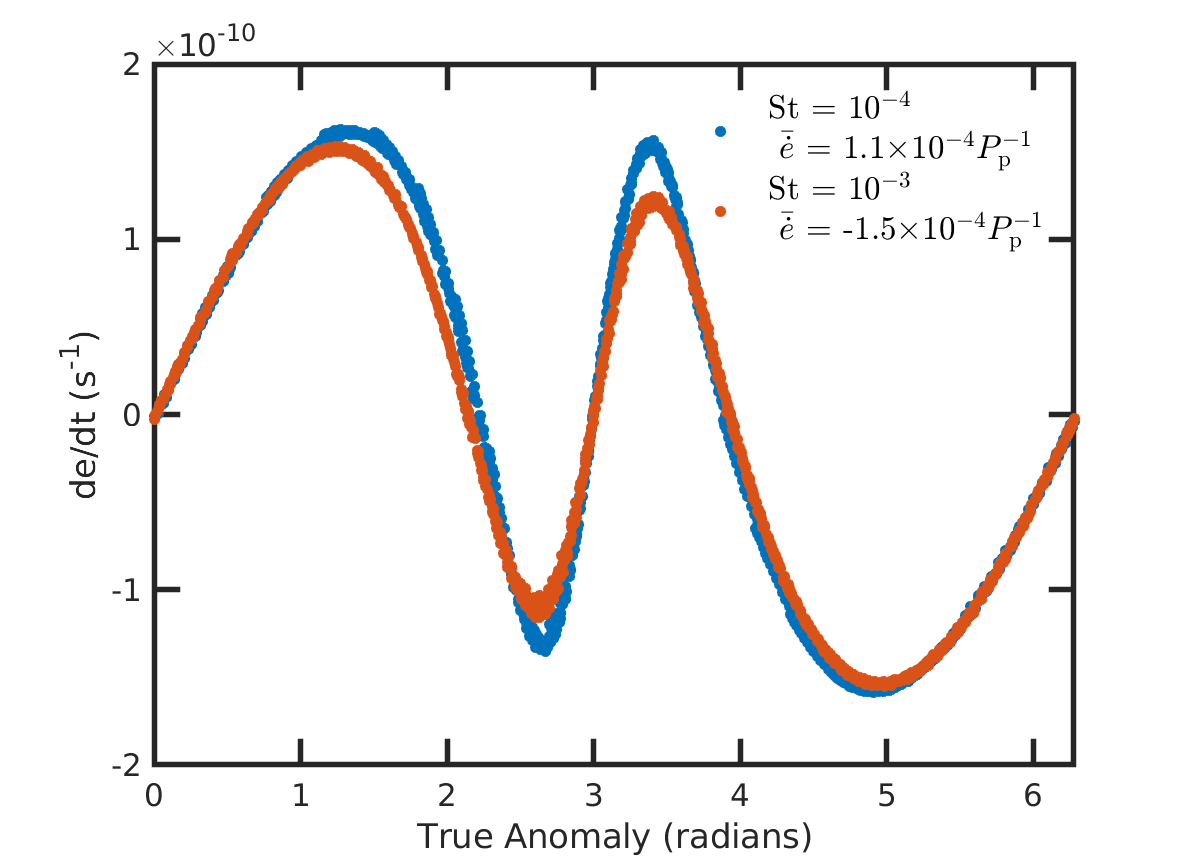}\\
    \caption{Values for $de/dt$ as a function of the planet's true anomaly between the times of 100,000 to 101,000 binary orbits. Blue points refer to the planet that got ejected, with red points being for the planet that remained stable. The legend shows the average change in eccentricity per planetary orbit, averaged over 10 planetary orbits.}
    \label{fig:eject_dedt}
\end{figure}

With fig. \ref{fig:eject_ecc} showing that the contribution from the disc is increasing the ejected planet's eccentricity, we now show the contributions from the disc to the eccentricity evolution as a function of the planet's true anomaly.
Figure \ref{fig:eject_dedt} shows this for a period of 1000 binary orbits for the planet that remained on a stable orbit (red points) and for the ejected planet (blue points).
With the planets' orbits being aligned with the disc, the effect of the higher density of gas and dust at the apocentre can be easily seen, both in magnitude and direction.
Either side of the planets' apocentres, the disc acts to increase and decrease the eccentricity when the planets are moving away from or towards it, as expected from the torques that are exerted during these phases. It is noteworthy that the differences between the two runs are small, indicating that it is subtle differences in disc structure and planetary orbital elements that lead to their different qualitative evolution rather than obvious systematic differences.
The legend in fig. \ref{fig:eject_dedt} shows the averaged values for $de/dt$ taken over 10 planetary orbits.
For the stable case, the average $de/dt$ is equal to $\sim -1.5\times10^{-4}$ per planetary orbit, a gradual decrease in eccentricity.
However, for the ejected case $de/dt \sim10^{-4}$ per planetary orbit on average, showing the gradual \emph{increase} in eccentricity that ultimately led to the planet being ejected.

\section{Discussion and Conclusions}
\label{sec:conclusions}

We have used 2D two-fluid hydrodynamical simulations to examine the effects that dust with different Stokes numbers and abundances has on the evolution and structure of the inner regions of circumbinary discs, and on the migration of planets in close binary systems that are analogues of Kepler-16 and Kepler-34. We have considered Stoke numbers $10^{-4} \le \Stokes \le 10^{-1}$ and \dtg ratios of 0.01 or 1.

\subsection{Effects of dust on disc structure}
Initially we examined the evolution of circumbinary discs without an orbiting planet. We allowed the gas and dust in the disc to evolve and settle into a quasi-equilibrium state, as a true equilibrium state cannot exist when dust can drift inwards from large radii. When comparing the evolution of discs with low \dtg ratios to a disc with no dust, we found only minor differences in the structure of the disc as measured by the cavity size and eccentricity, and the radial profiles of surface density and eccentricity.

For discs with \dtg ratios of unity, significant differences arose in cavity sizes and eccentricities as a function of the Stokes number. For $10^{-4} \le \Stokes \le 10^{-3}$, corresponding to strongly coupled dust, the cavity sizes and eccentricities were found to be reduced only modestly compared to a dust-free disc. Increasing $\Stokes$ to $10^{-2}$ and $10^{-1}$ had the effect of progressively circularising the cavity and reducing its size.
For $\Stokes=10^{-1}$ the cavity size reduced to become comparable to the dynamical stability limit for test particles\citep{Holman99}, with the eccentricities becoming very small.

The discs with \dtg ratios of unity and $10^{-2} \le \Stokes \le 10^{-1}$ the presence of a large density and pressure bump at the edge of the cavity seen in other studies \citep[e.g.][]{Mutter17D,Thun17} was diminished as the dust drift caused the gas surface density profiles to flatten, especially for the larger of these Stokes numbers. Over the run times we considered, all runs with $\Stokes=10^{-1}$ resulted in a
ring of dust accumulating at the cavity edge, increasing the \dtg there by a factor of 20 compared to its initial value.

\subsection{Migration behaviour and stopping locations}
To examine the stopping locations of migrating planets in the circumbinary discs, we placed $20\me$ planets in circular orbits far outside of the cavity region.

In agreement with previous works \citep[e.g.][]{Pierens13,Mutter17P,Thun18}, planets in the discs with low \dtg ratios migrated to the inner cavity region, before stalling there. 
For both the Kepler-16 and -34 analogues, their final locations were more distant from the central binary than the observed values for Kepler-16b and Kepler-34b, although we note that the planet masses considered in this study do not correspond to the observed planet masses. There was also very little change in the migration behaviour and stopping location of the planets with respect to the Stokes number.
The only noticeable difference was in the runs with the largest Stokes number, where the planets were able to migrate slightly closer to the central stars because the build-up of dust allowed the interactions between planets and discs to become nonlinear, resulting in the planets forming their own gaps and migrating closer to the stars.

For the discs with \dtg ratios of unity, the Stokes number heavily influenced the evolution of the planets' orbits.
Differences in behaviour were also observed across the two systems examined, Kepler-16 and -34, as a result of the planets being able to open gaps in the dust discs around Kepler-16.
Around both systems, the planets in discs with small Stokes numbers migrated into the cavity region before stalling at the cavity, as with the planets in the low \dtg ratio discs described above.
Around Kepler-34, the planets in the discs with larger Stokes numbers underwent runaway migration and stopped at the cavity edge, where they formed a gap and migrated closer to the central binary towards, eventually stalling quite close to the observed location of Kepler-34b. In the case with $\Stokes=10^{-1}$, the small size of the cavity caused the planet's final stopping location to be at the stability limit.

For the planets in the discs with \dtg of unity and $\Stokes =10^{-1}$ in the Kepler-16 analogues, migration was different because planets were able to open a deep gap in the dust disc, and this caused the inward migration to occur at a very slow rate.

In some cases we found that planets could be ejected from the system. In each case, this occurred because a secular resonance between the precessing disc and planet arose, and this drove the planet eccentricities to a large value such that they then scattered off the central binary.

\subsection{Future work}
In this paper we have presented a suite of proof-of-concept simulations that demonstrate that the inwards drift and accumulation of dust particles at the edge of the inner cavity in a circumbinary disc can change the structure of the cavity and the parking locations of migrating planets. However, many of the ingredients of our models are highly simplified and will be improved in future work. In a forthcoming publication we will present simulations where the dust fluid no longer has a fixed Stokes number but instead corresponds to a fixed particle size. In a real disc, however, the dust has a continuous range of particle sizes, and the size distribution may change due to growth and fragmentation, and this is something we will explore in future models. We will also move away from using a locally isothermal equation of state and improve the treatment of the disc thermodynamics by incorporating irradiation of the disc and radiation transfer, and adopting 3D instead of 2D models will be an important improvement. The development of a sophisticated model will allow us to use the observed diversity of orbital parameters for circumbinary planets as a testing ground for modelling the dynamics of protoplanetary discs and planet migration.

\section*{Data Availability}
The data underlying this article will be shared on reasonable request to the corresponding author.

\section*{Acknowledgements}
We thank the anonymous referee for useful comments on the paper.
GALC was funded by the Leverhulme Trust through grant number RPG-2018-418. RPN acknowledges support from STFC through grants ST/P000592/1 and ST/T000341/1.
This research utilised Queen Mary's Apocrita HPC facility, supported by QMUL Research-IT (http://doi.org/10.5281/zenodo.438045). This work was performed using the DiRAC Data Intensive service at Leicester, operated by the University of Leicester IT Services, which forms part of the STFC DiRAC HPC Facility (www.dirac.ac.uk). The equipment was funded by BEIS capital funding via STFC capital grants ST/K000373/1 and ST/R002363/1 and STFC DiRAC Operations grant ST/R001014/1. DiRAC is part of the National e-Infrastructure.
This research received funding from the European Research Council (ERC) under the European Union's Horizon 2020 research and innovation programme (grant agreement n$^\circ$ 803193/BEBOP)

\bibliographystyle{mnras}
\bibliography{references}{}

\appendix

\section{High resolution ejection figure}
\label{sec:app_eject}
\begin{figure}
    \centering
    \includegraphics[scale=0.6]{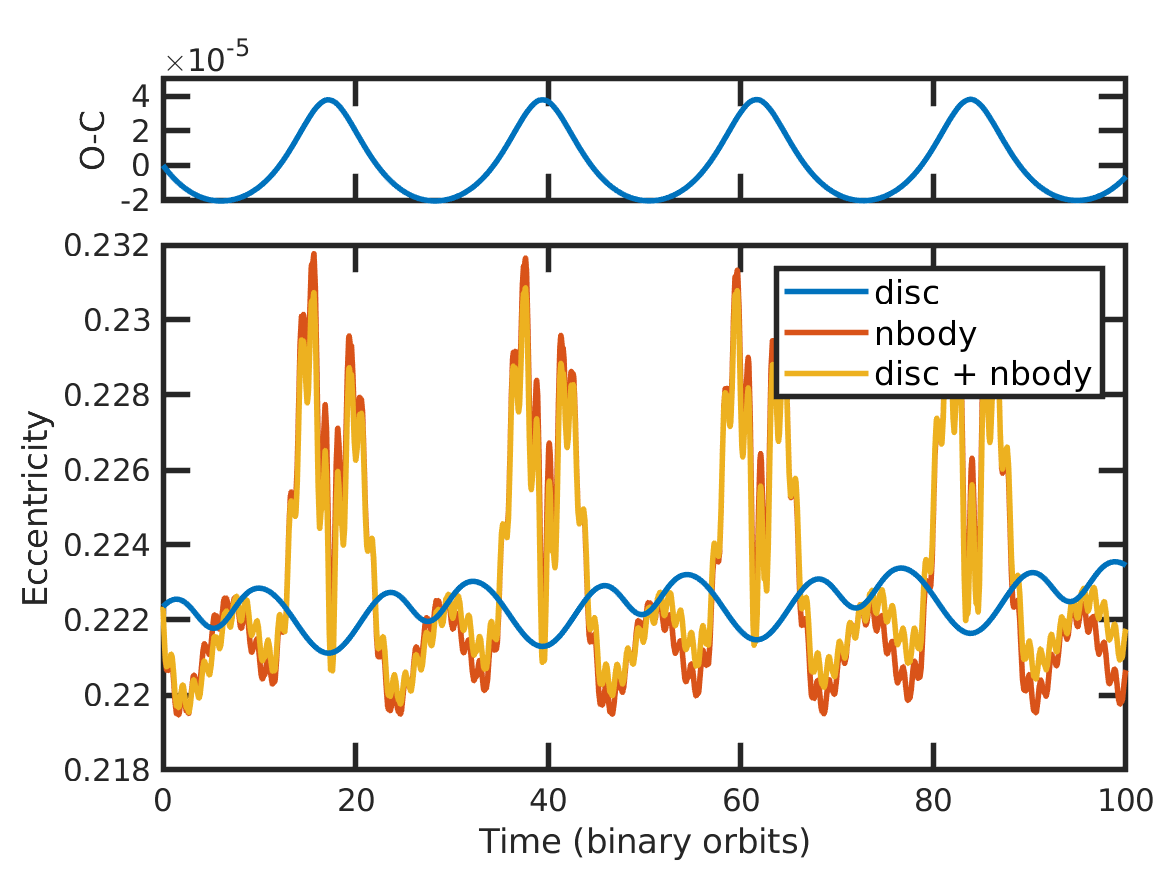}\\
    \caption{Time evolution of planet eccentricities for the ejected planet with high time resolution. The colours refer to the eccentricities where only N-body (red), disc (blue), disc and N-body (yellow), effects are included. The top panel shows the residuals between the observed eccentricity and the calculated eccentricity including the disc and N-body effects.}
    \label{fig:eject_appendix}
\end{figure}

Figure~\ref{fig:eject_appendix} shows the results of integrating the eccentricity of the ejected planet using high temporal resolution data outputs of the forces acting on the planet from the hydrodynamical simulation described in Sect.~\ref{sec:ejections}. The outputted times occur in the time interval 100,000--100,100 binary orbits shown in fig. \ref{fig:eject_ecc}, with the initial conditions for the disc and N-body positions/velocities taken from those recorded at 100,000 binary orbits.

\label{lastpage}
\end{document}